\documentclass[usenatbib]{mnras}
\usepackage[T1]{fontenc}
\usepackage{amsmath}
\usepackage{multirow}
\usepackage{graphicx}
\providecommand{\tabularnewline}{\\}
\begin{document}
\title{The Extended Solar Cycle and Asimmetry of the Large-Scale Magnetic Field}
\author[V.~Obridko et al.]{V.N.~Obridko$^{1,2}$\thanks{obridko@izmiran.ru}, A.S.~Shibalova$^{1}$, D.D.~Sokoloff$^{1,3,4}$\thanks{sokoloff.dd@gmail.com}, \\
$^{1}$IZMIRAN, 4, Kaluzhskoe  Shosse, Troitsk, Moscow, 108840, Russia\\
$^{2}$Central Astronomical Observatory of the Russian Academy of Sciences at Pulkovo, St.Petersburg, Russia\\
$^{3}$Department of Physics, Lomonosov Moscow State University, Moscow, 119991, Russia\\
$^{4}$Moscow Center of Fundamental and Applied Mathematics, Moscow,
119991, Russia}

\date{Received ..... ; accepted .....}
\maketitle

\begin{abstract}
Traditionally, the solar activity cycle is thought as an interplay of the main dipole component of the solar poloidal magnetic field and the toroidal magnetic field. However, the real picture as presented in the extended solar--cycle models is much more complicated. Here, we develop the concept of the extended solar cycle clarifying what zonal harmonics are responsible for the equatorward and polarward propagating features in the surface activity tracers. We arrive at a conclusion that the zonal harmonics with $l=5$ play a crucial role in separating the phenomena of both types, which are associated with the odd zonal harmonics. Another objective of our analysis is the role of even zonal harmonics, which prove to be rather associated  with the North-South asymmetry of the solar activity than with its 11-year solar periodicity.
\end{abstract}


\section{Introduction}

The basic modern concept of origin of the solar magnetic field is based on the dynamo theory, which seems to be a fairly developed area of physics. The concept sounds as follows. The initial poloidal magnetic field being affected by differential rotation produces a toroidal magnetic field, which, in turn, being affected by the meridional circulation and/or Babcock-Leighton mechanism, restores the poloidal magnetic field near the poles.  At the time of the magnetic field reversal, the equatorial magnetic field becomes minimum, while the polar one becomes maximum. This is how the 11-year cycle of magnetic energy variation arises. The link between two successive cycles is due to the fact that the restored polar magnetic field, which is the origin of the following cycle, has the sign opposite to the field sign in the previous cycle. As a result, the sign of the toroidal magnetic field in the $n$-th cycle is opposite to that in the $(n-1)$-th cycle, which is reflected in the Hale polarity law. So, the real physical period of the solar activity cycle is not 11, but 22 years.

This clear and straightforward scheme has to incorporate two effects, which somehow complicate it and modify the relation between the cycles. First of all, the concept of solar activity has to include  the “Extended Solar Cycle".  The term appeared in literature in 1988 \citep{A88, Wetal88}, although during that decade, observational evidence appeared indicating that the magnetic activity of one cycle overlapped for some period of time (often up to several years) with that of the previous cycle \citep{LN83}. As observed on the surface, the extended solar cycle starts during the sunspot maximum at high latitudes and consists of a relatively short polarward branch (described as "rush to the poles") and a long equatorward branch, which continues through the solar minimum and the following sunspot cycle \citep{KPG21}, see also \cite{MIetal21} for review.

The culmination of many years of painstaking observations, cataloging, and individual publications by a number of prominent observers of that time was the idea of the extended cycle as a pattern formed by a host of observables in latitude and time. Some of those observables are easily associated with magnetism in nature, prominences and filaments \citep[e.g.][]{B33, HH75, MI92, Tetal16} and ephemeral active regions \citep[e.g.][]{HM73}, as well as the gross features in the Sun's green-line corona \citep[e.g.][]{A97}, while the others, like the zonal patterns of the torsional oscillation \cite[e.g.][]{HL80, SW87, W94}, still need explanation. It is also interesting to note papers that refer to the long baseline of geomagnetic data  consistent with the signs of temporally overlapping activity cycles on the Sun \citep{M75, LS81}.

\cite{MIetal20} applied (discrete) Hilbert transforms to more than 270--year series of monthly sunspot numbers to identify the so-called "termination" events that mark the end of the previous 11-yr sunspot cycle, the enhancement/acceleration of the present cycle, and the end of the 22-yr magnetic activity cycle. In fact, we will show below that sometimes three cycles may take place on the Sun simultaneously instead of one. The cycles overlap at the equatorial, middle and high latitudes (see Sec. 3).

One more effect to be incorporated in the dynamo concept is the North-South Asymmetry of solar activity as a whole and the corresponding magnetic field asymmetry, in particular. Although the North-South Asymmetry  of solar activity has been known for a long time, the attempts to separate theoretical modeling of the northern and southern hemispheres and to estimate the difference between them appeared only 10–20 yr ago. Among the possible mechanisms responsible for differences in the activity characteristics between the northern and southern hemispheres, different authors mentioned the stochastic effect of the convection \cite{Hetal94}, the counter effect of the generated magnetic field on matter flows, which is described by nonlinear dynamo equations \citep{W10}, and, simply, the existence of a primary relic field \citep{BL84, M07}. The latter, however, is difficult to reconcile with a strong variability of the asymmetry both in sign and in absolute value on short and long timescales. \cite{Betal13}  show that an $\alpha - \Omega$--type solar dynamo normally operates independently in two hemispheres --– if the dynamo dies in one   hemisphere due to subcritical dynamo number, the dynamo in the other hemisphere operates without being much affected. Of particular interest are the studies that describe the counter effect of the magnetic field on the differential rotation characteristics \citep{SNR94, T97}. This negative correlation is confirmed in general in a number of publications \citep{HW90, KN90, OS00b, OS00a, OS16}. These results impose additional restrictions on the problem of the asymmetry modeling on the basis of the dynamo mechanism. A detailed review of the mechanisms of interaction of hemispheres both from the experimental and from the theoretical points of view is given in \cite{Netal14}.

Many authors have investigated the asymmetry using various indices of solar activity, such as the indices of sunspot activity, solar flares, filaments, prominences, radio and gamma bursts, coronal radiation, and solar magnetic field. The state of the art of the problem is discussed by \cite{VB90, Cetal93, Cetal07, Letal02, Metal02, Tetal06, SR10}, as well as in our papers \cite{Betal05, Betal08, BO11}. \cite{B11}  considered such a manifestation of the North-South Asymmetry as the asymmetry of the monthly mean sunspot latitudes (i.e., the sunspot production centers) in the northern and southern hemispheres, and its relationship with the commonly analyzed asymmetry of total sunspot areas. The behavior of these parameters suggests a disbalance between the two solar hemispheres, which is most clearly pronounced in the epochs of the cycle minimum.

\cite{BO17} demonstrated that the asymmetry sign provides information on the behavior of the asymmetry. In particular, it displays quasi-periodic variation with a period of 12 yr and quasi-biennial oscillations of the asymmetry itself. The statistics of the so-called monochrome intervals (long periods of positive or negative asymmetry) are considered and it is shown that the distribution of these intervals is described by the random distribution law. This means that the dynamo mechanisms governing the cyclic variation of solar activity must involve random processes. At the same time, the asymmetry modulus has completely different statistical properties and is probably associated with processes that determine the amplitude of the cycle. One can reliably isolate an 11-yr cycle in the behavior of the asymmetry absolute value shifted by half a period with respect to the Wolf numbers. It is shown that the asymmetry modulus has a significant prognostic value: the higher the maximum of the asymmetry modulus, the lower the following Wolf number maximum.

Both effects (extended cycle and asymmetry) are intrinsically associated with the relation between various magnetic field components, which, in turn, depend on the spatial scales, latitudes, and longitudes. In particular, the overlapping of cycles is connected with the odd zonal harmonics, while the asymmetry involves even harmonics. Having a reliable long-term database of the solar magnetic field we can apply the above analysis and obtain the above scales, as well as their temporal evolution. We describe below the database and the tools to extract the mentioned scales (Sect.~\ref{data}).

\section{Database, Basic Equations, and Calculation Method}
\label{data}
The main data series used for analysiswere  obtained at the John Wilcox Stanford Observatory (WSO). These series start May 1976 (Carrington Rotation 1641)
and  continue till the present (http://wso.stanford.edu/forms/prsyn.html). We have decomposed the surface field observed at WSO into its harmonic components and presented the time evolution of the mode coefficients for the past four sunspot cycles. We have been working with the WSO synoptic maps of the light-of-sight photospheric magnetic field component  converted into a sum of associated Legendre polynomials $P_{ml}$ \citep{Betal20}.

All components of the magnetic field at any point of a spherical layer from the photosphere to the so-called source surface can be calculated under potential approximation from the line-of-sight field observations. The source surface is by definition a spherical surface, where all field lines are radial. It is suggested to lie at a distance of $R_s = 2,5 R_o$ from the center of the Sun.

The equations for calculating the magnetic field components are written as
follows:

\begin{eqnarray}
B_r= \sum_{l,n,m} P_l^m (\cos \theta ) (g_l^m \cos m \phi + h_l^m \sin m \phi)\times \\
\nonumber \times ((l+1) (R_0/R)^{l+1} - l (R/R_s)^{l+1} c_l),\\
B_\theta = -\sum_{l,n,m} {\frac{\partial P_l^m (\cos \theta)}{\partial \theta}} (g_l^m \cos m \phi + h_l^m \sin m \phi) \times \\
\nonumber ((R_0/R)^{l+2} + (R/R_s)^{l-1} c_l),\\
B_\phi = -\sum_{l,n,m} \frac{m}{\sin \theta } P_l^m ( \cos \theta ) (h_l^m \cos m \phi - g_l^m \sin m \phi) \times \\
\nonumber \times ((R_0 / R)^{l+2} + (R / R_s)^{l-1} c_l).
\label{Gauss}
\end{eqnarray}
Here, $0 \le m, l < N $ (usually, $N \le 9$), $c_l=-(R_0/R_s)^{l+2}$, $P^{m}_{l}$ are the Legendre polynomials, and $g_l^m$, $h_l^m$ are the harmonic coefficients. The latter was calculated by ourselves from WSO Stanford data. To find the harmonic coefficients, $g_l^m$ and $h_l^m$, and thus, to fully determine the solution, we had to use the following boundary conditions. By the lower boundary condition we usually adopted the synoptic map, constructed from observations of the line-of-sight field component in the photosphere ($r<R_{\rm phot}$). The upper boundary is the source surface, where all field lines are radial.

In further calculations, one can follow two different ways: to proceed from orthogonality
of the Legendre polynomials \citep{HS86}, or to use the least-squares method \citep{KI94}. The former method provides a straightforward physical interpretation of the coefficients and directly relates them to various current and field systems in the solar atmosphere. The advantage of this method is that every coefficient is calculated independently, and the ultimate number of terms in the sums of Equations (1) and (2) can be increased without re-calculating the formerly calculated coefficients. The least-squares method provides a much better approximation of the photospheric field in the case, when the number of measured points in the synoptic map is large enough. The method is more convenient for computer calculations, than the former one, and it can be readily modified for different metrics. On the other hand, in case of changing the order of the system, or the form of data, the representation requires re-calculation of all coefficients.

It is not obvious that the best approximation achieved in the photosphere, will be valid at other levels, e.g. at the source surface. Of course, both methods yield strictly coinciding results at N tending to infinity; however at any finite $N$  \citep[in][$N \le 9$]{HS86, KI94}, the similarity of the obtained results is not convincing and should be verified.

\section{Odd axisymmetric harmonics and overlapping phase}

Now, following \cite{OY89} (see also \cite{OS92}), we can calculate the mean
square radial component of the magnetic field on the sphere of radius $R$. Using the orthogonality of polynomials,
we can perform integration over a spherical surface analytically and obtain the result in closed form. As an
illustration, we give formulas for two boundary surfaces, i.e.  the photosphere $i(B_r)_{|_{R_0}}$ and the source
surface $i(B_r)_{|_{R_s}}$ .

\begin{equation}
i(B_r)_{|_{R_0}}=\sum\limits_{lm}\frac{(l+1+l\zeta^{2l+1})^2}{2l+1}(g^2_{lm}+h^2_{lm}),
\end{equation}

\begin{equation}
i(B_r)_{|_{R_s}}=\sum\limits_{lm}(2l+1)\zeta^{2l+4}(g_{lm}^2+h_{lm}^2),
\end{equation}
where $\zeta=R_0/R_s$. This means that the contribution of the $l^{th}$ mode to the mean magnetic field contains an $l$-dependent coefficient. In these formulas, $i(B_r)_{|_{R_0}}$
and $i(B_r)_{|_{R_s}}$ are the mean square radial components of the magnetic field in the photosphere and at the source surface, respectively. In our calculations, the radius of the source surface $R_s=2.5R_0$ and, thus, $\zeta$ = 0.4.

\begin{figure}
\includegraphics[width=0.97\columnwidth]{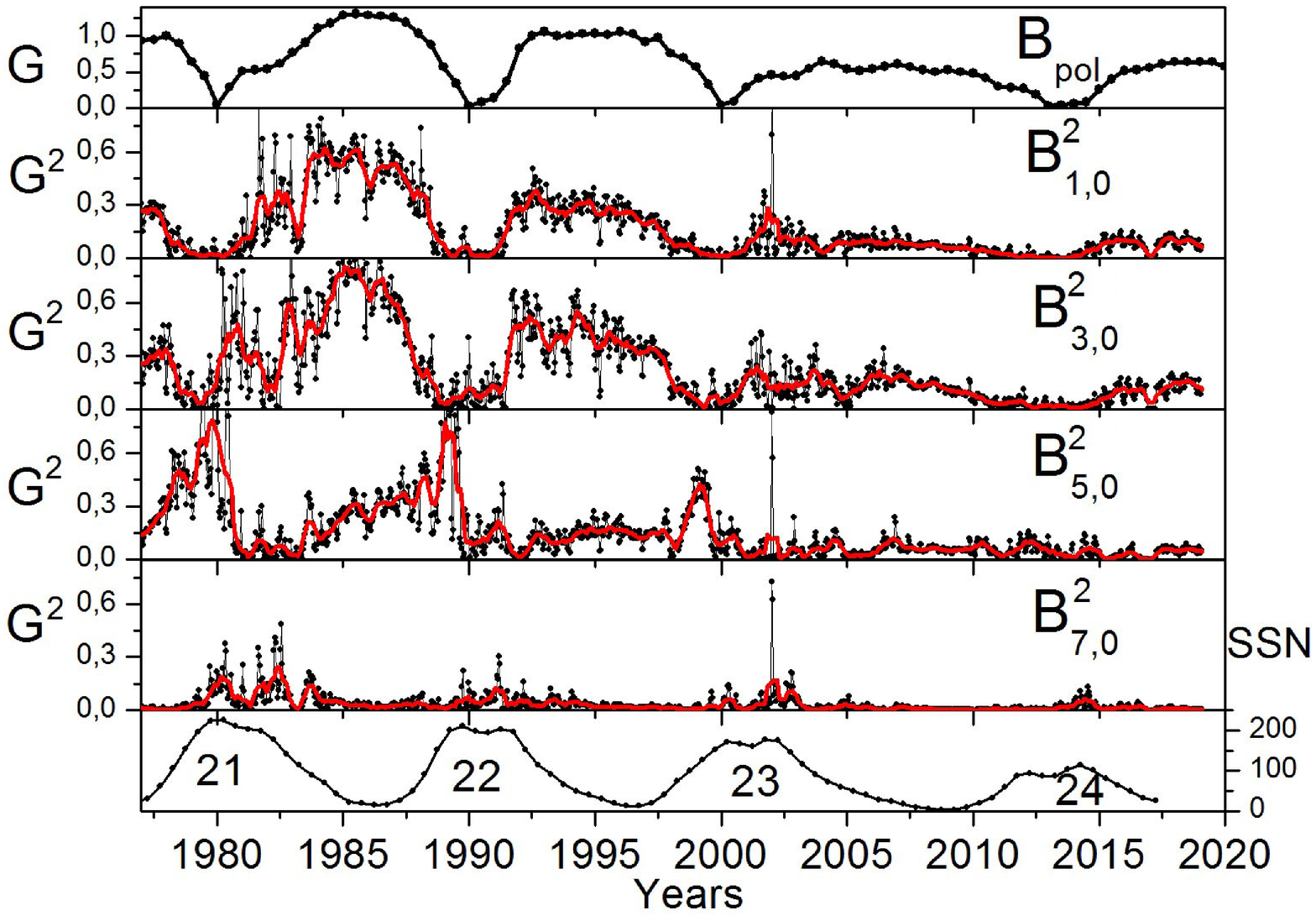}
\includegraphics[width=0.90\columnwidth]{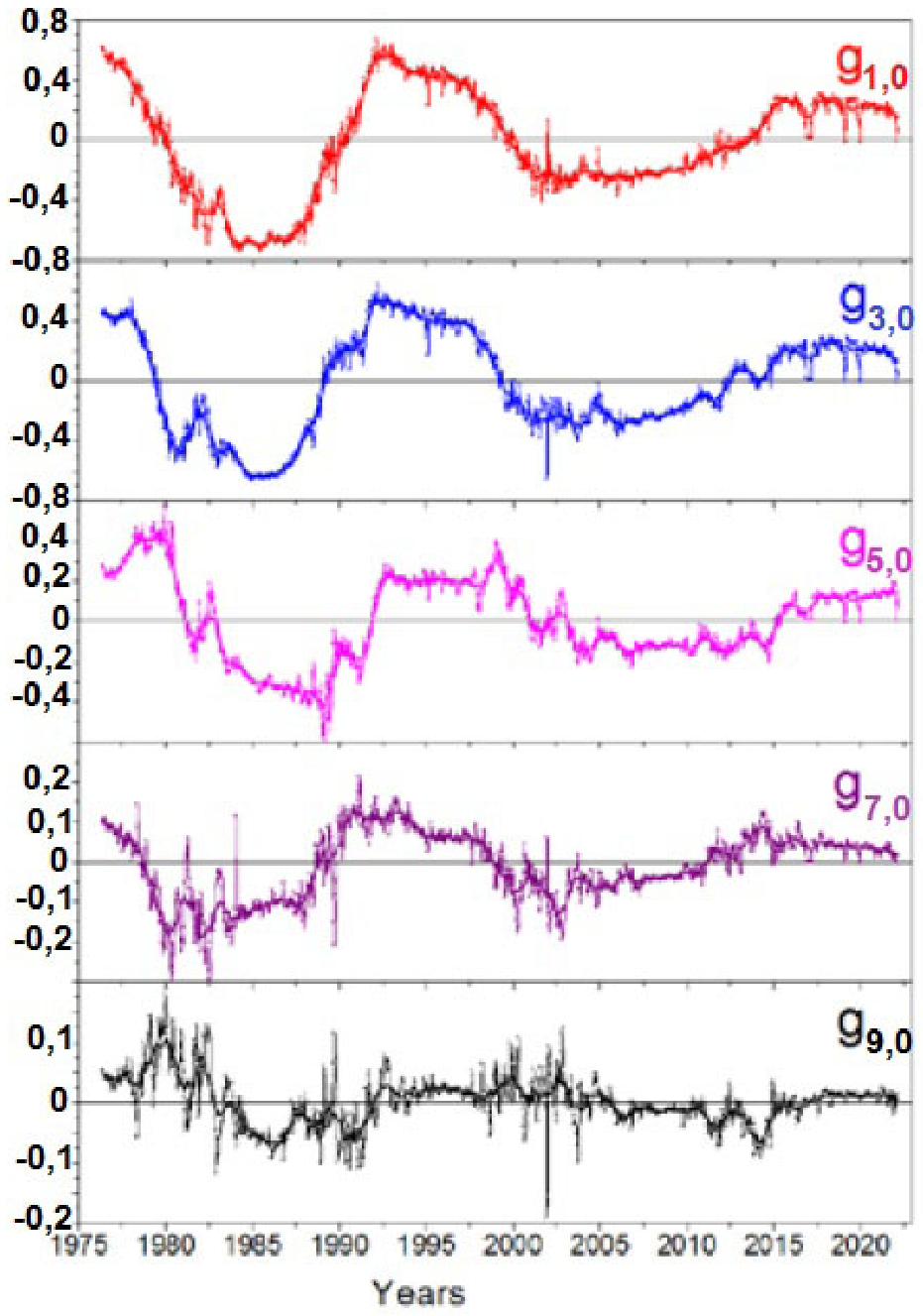}
\caption{The upper panel shows the time dependence of squared magnetic field connected with the first odd axisymmetric harmonics up to $l=7$ \citep{Obetal21}. The lowest graph represents the time variation
of sunspot numbers (SSN) with the cycle numbers indicated; the top graph is the polar field given after http://wso.stanford.edu/Polar.html (evolution of the absolute value of the polar magnetic field is shown). The lowest panel shows amplitudes of zonal harmonics $l=1,3,5,7,9$. {\bf }}
\label{f1}
\end{figure}

Specifying $l$ and $m$ in the above equations, we can calculate the square of the mean magnetic field connected with contribution of a certain component. Fig.~\ref{f1} upper panel illustrates the time dependence of the first  odd axisymmetric harmonics up to $l=7$. The bottom graph shows the time variation of sunspot numbers; the top graph is the polar field given after http://wso.stanford.edu/Polar.html).

The representation in the form $i(B_r)_{|_{R_0}}$ and $i(B_r)_{|_{R_s}}$ is convenient, since it characterizes the energy of the contribution of various components of the magnetic field. However, when calculating the structure, it is necessary to use the amplitudes harmonics. These amplitudes for odd zonal harmonics are shown in the bottom panel of Figure 1 and for even harmonics in Figure 4. All values along the y-axis are given in units of G.

Fig.~\ref{f1} confirms that the harmonics with $l=1$ and $l=3$ behave directly as expected from the solar dynamo theory, i.e. they follow the evolution of the polar magnetic field  and are in antiphase with the sunspot number variation, which is believed to be related to the evolution of the toroidal magnetic field.  On the contrary, the harmonics $l=5$
behave quite differently and become maximum just at the beginning of the minimum of the polar magnetic field (\cite{Obetal21}).

The evolution of the $l=5$ mode becomes clear when compared with the time-latitudinal diagram of the large-scale solar magnetic field (Fig.~2, upper panel). This map is constructed using the  Kitt Peak observation data obtained with a good ground--based resolution. Long-term plots of the solar field are generally called ‘supersynoptic’ maps rather than ‘synoptic’ maps, which generally refer to single Carrington Rotations.

\begin{figure*}
\includegraphics[width=2\columnwidth]{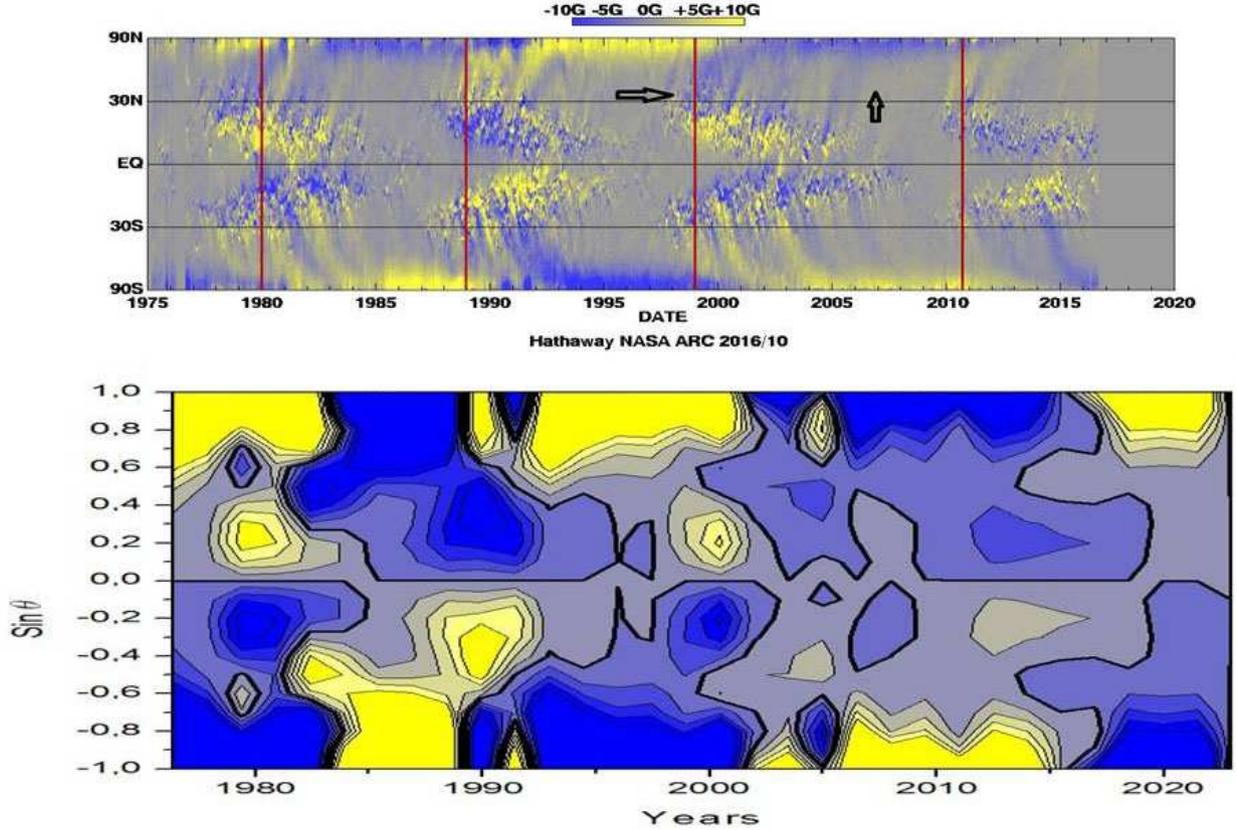}
\caption{Magnetic data (the yellow elements correspond to the positive polarity; the blue
ones correspond  to the negative polarity). The upper panel shows a supersynoptic map based on Kitt Peak data. The horizontal lines represent co-latitude $\theta$ in units of $\sin \theta$. The horizontal arrow shows the beginning of RTP in 1999, while the vertical arrow shows the Ripples region in 2007. The vertical red lines show the maxima of the fifth harmonic (cf. Fig.~1). The lower panel shows the supersynoptic map  based the WSO data as a sum of zonal ($m=0$) harmonics with $l=1, 3, 5, 7, 9$ calculated by Eq.~(1). }
    \label{f2}
\end{figure*}

This quite a complicated supersynoptic map of the magnetic field makes it possible to identify manifestations of the following processes.  We see there a region where the flux migrates polewards from
about $-35^o$ and $+35^o$ in the southern and northern hemispheres, respectively. This phenomenon usually called Rush-to-the-Poles (RTP) is mainly associated with large-scale magnetic fields. One of such waves is marked with a horizontal arrow in the vicinity of 1999 on the upper panel in Fig.~\ref{f2}.  Near this date (1999) we see a yellow band (that is, the positive polarity of the magnetic field), which came to the pole from the middle latitudes. At the same time, a blue band (that is, of negative polarity) has already appeared in the middle latitudes, which will drift towards the pole and reach it only in 2010. And at the same time, there is already a third band of intermittent color, but predominantly yellow (that is, the same as on the pole). This wave is associated with local fields and is drifting towards the equator and will reach it by 2005. After RTP  reaches the pole, it replaces the previous activity wave, which results in the reversal of the magnetic field.

Of course, the map contains also the well-known wave propagating from the middle latitudes to the equator. The wave manifests itself in the form of a standard Maunder butterfly diagram and is mainly associated with local magnetic fields of active regions. Both the poleward and the equatorward waves appear almost simultaneously and have opposite dominant polarities of the magnetic field.

The point, however, is that in certain time intervals, the map reveals two Rush-to-the-Poles waves with opposite magnetic field polarities. As one of them nearly reaches the pole, the other just appears at mid latitudes. In these time intervals, the map contains three regions of magnetic field polarities opposite in a given zone with respect to the neighbouring one.

We emphasize that the polar wave contains the regions where the integral contributions of the magnetic fields of both polarities are more or less equal in absolute value. This is clearly visible on the map based on KItt Peak data. As for the lower resolution  WSO  magnetograph data, the equatorward wave is less pronounced (see, e.g., the diagram presented at http://wso.stanford.edu/gifs/all.gif -- the smoothed solar zonal field for 4 cycles, which we do not reproduce here). In some cycles, the equatorward wave is almost unrecognizable.

In addition, it should be noted that the method of reconstruction of the magnetic field from observation data contains some debatable points, such as the field potentiality at the photosphere surface, the drop of accuracy for the Gauss coefficients because of averaging of synoptic maps over a Carrington rotation, etc. In principle, the results based on WSO and Kitt Peak data  might have been substantially different; however, fortunately we see from Fig.~2 that the results are in a reasonable agreement in all substantial details taking into account different resolution of observations.

Another less investigated phenomenon is that of thin strips migrating from the solar equator to the poles (marked by a vertical arrow on the map). The strips appear mainly in the decay phase of the 11-year activity cycle. Their lifetime is about 1--3 years. \cite{Vetal12, UT13} called them «Ripples». We shall discuss this phenomenon later, in Sect.~4.

A simultaneous presence of two Rush-to-the-Poles events with opposite magnetic field polarities on the Sun called the attention of experts. In particular, \cite{MIetal14, MIetal19, MIetal20, MIetal21} introduced the concept of cycle terminator as an instant when two activity waves depart from the latitude of $55^o$, propagating one equatorwards and the other, polewards.  We extend this concept and discuss the  propagation of three coexisting activity waves, which does not occur instantly, but over a time interval of about a year.
 We propose to call this time interval ''the overlapping phase''. As seen below, the overlapping phase can be quantitatively described in terms of the 5th zonal magnetic field harmonics,  which, in this connection, can be referred to as the height of the overlapping phase.

Let us now  clarify the difference between the RTP and equatorward wave. We expect that a four-cycle diagram of the radial magnetic field plotted with odd harmonics calculated by Eq.~(1) may be instructive here. Fig.~3 (upper panel) shows a synoptic map obtained using three first odd zonal harmonics ($l=1, 3, 5$, $m=0$). It is evident that all RTP phenomena are present and the cycle overlapping is well pronounced, while the equatorward wave is almost invisible.

Using the Hilbert transform, \cite{MIetal20} calculated the terminator instants for 140 years with the accuracy of 0.01 year. In particular, after 1978, the terminators occurred in 1978.00, 1988.25, 1998.25, and 2011.08. The data are quite close to our overlapping phase times. \cite{MIetal20} propose to use the time between two successive terminators as a prognostic index for the amplitude of the solar cycle and expect that Cycle 25 will be very high. However, the idea requires that the terminators be determined with a very high accuracy.

Fig.~\ref{f1} above shows a general decay of activity from cycle to cycle accompanied by a decrease in the amplitude of the $l=5$ harmonics in the overlapping phase. Of course, four cycles do not provide sufficient statistics for a reliable correlation analysis; however it looks plausible that the following cycle would be rather low. Note that a reliable estimate of the amplitude of the  $l=5$ harmonic can be based on magnetic field data for several solar rotations or even for a single  rotation. E.g., back in early 2011 we could expect that Cycle 24  would be low.
 Similarly, we can claim now with the same degree of confidence that there are no reasons to believe that Cycle 25 may be stronger than Cycle 24  (see https://www.swpc.noaa.gov/products/solar-cycle-progression).

\begin{figure}
\includegraphics[width=1\columnwidth]{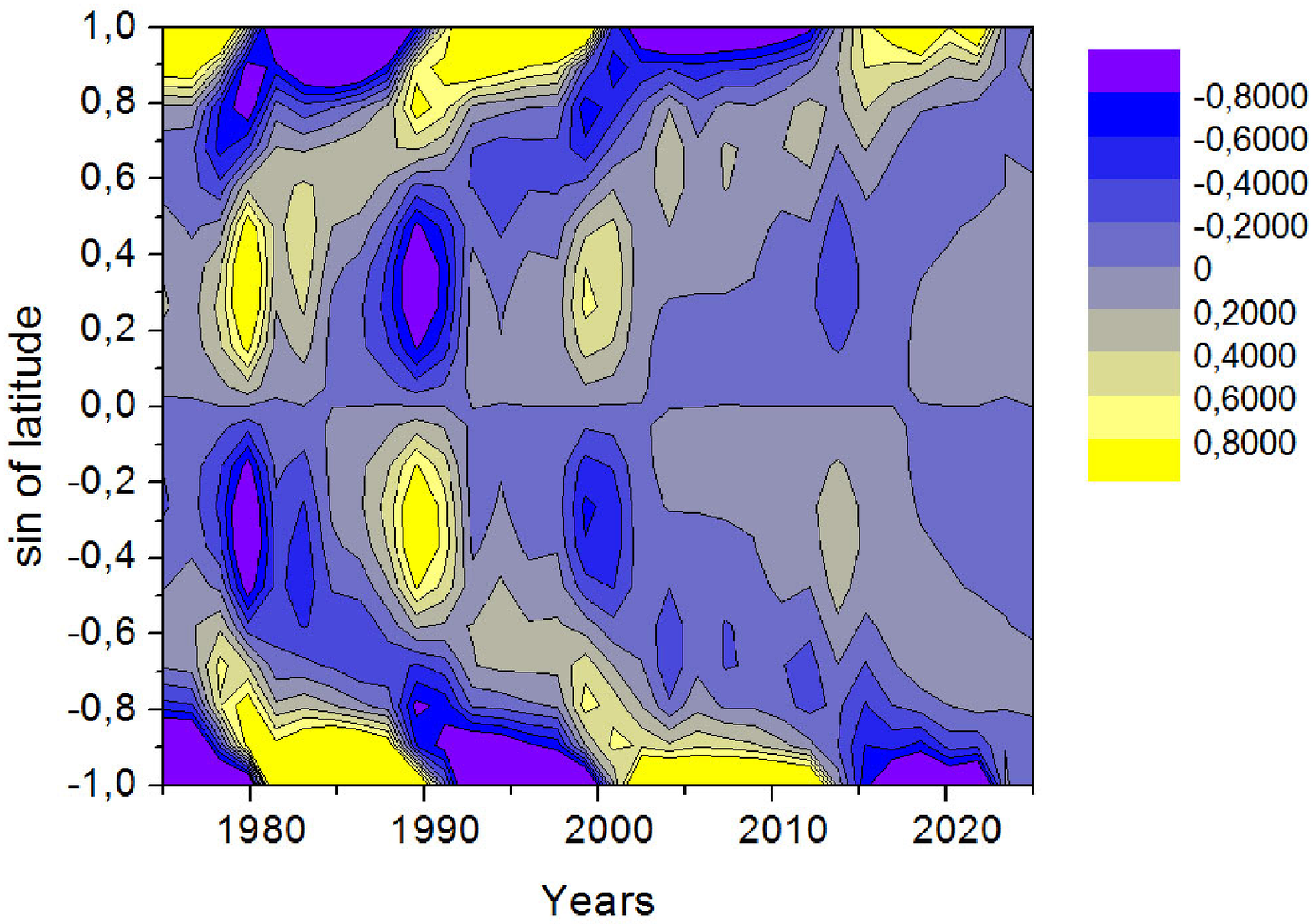}
\includegraphics[width=1\columnwidth]{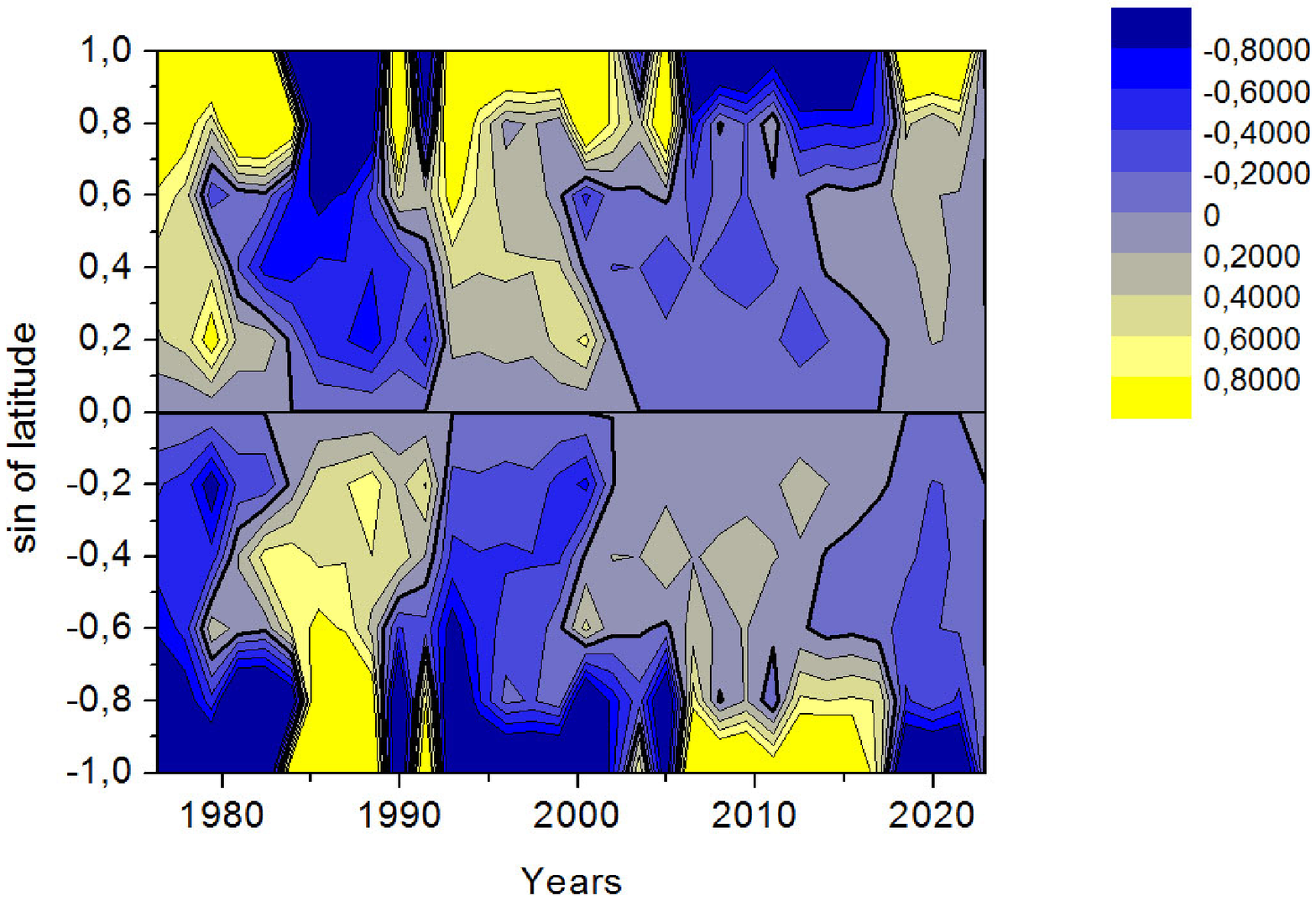}
\caption{Superynoptic map of the large-scale radial magnetic field reconstructed from three first odd zonal harmonics ($l=1,3,5$)smoothed over three Carrington rotations (upper panel) and from three  odd zonal harmonics ($l=5,7,9$) smoothed over three Carrington rotations (lower panel).}
    \label{f3}
\end{figure}

In Fig.~\ref{f3}, the upper panel shows a synoptic map of the large-scale radial magnetic field reconstructed from three first odd zonal harmonics ($l=1,3,5$) smoothed over three Carrington rotations. The map presents two poleward waves and basically agrees with a similar map produced directly from WSO observations (http://wso.stanford.edu/gifs/all.gif) of the antisymmetric total magnetic flux. The waves propagating equatorwards, however, are less pronounced on the plot, only some of their traces being visible. Perhaps, the reason is that we did not include higher-order zonal harmonics in the analysis or the waves were smoothed out. Still by directly including higher-order harmonics in the analysis and removing smoothing we do not restore the wave, but rather obtain a much more noisy map. Removing the $l=5$ harmonics,  we, on the contrary, obtain a map that does not reproduce any temporal drift at all.
For comparison, Fig.~\ref{f3} (lower panel) shows supersynoptic map of the radial magnetic field reconstructed from only three  odd zonal  harmonics ($l=5,7,9$) smoothed over three Carrington rotations. The map substantially differs from the previous one, because no Rush-to-the-Poles are visible, however the equatorward drift is well pronounced.

To sum up, we can say  that short before the reversal of the polar magnetic field, both large-scale and relatively small-scale magnetic fields appear virtually simultaneously at the photosphere surface. Having appeared almost simultaneously, the fields then begin to behave in quite a different way. The large-scale component, which is a unipolar entity, propagates producing  Rush-to-the-Pole effect. As a result, one can see two opposite in polarity Rush-to-the-Pole waves propagate simultaneously over the solar surface. In contrast, local magnetic fields are bipolar entities, which propagate equatorwards. Formally, the separation of waves of both types is quantified by harmonics with  $l=5$.

\section{Asymmetry and even zonal harmonics}

Now, let us discuss another phenomenon to be included in the initial oversimplified solar dynamo model, i.e. the North-South asymmetry. In Sect.~1, we mentioned some papers dealing with the study of the North-South asymmetry (NSA), but all of them are devoted to the analysis of structural elements associated with the local magnetic field data, i.e., with the toroidal field. It is reasonable, however, to expect that the asymmetry is a fundamental property, which is inherent of the large-scale magnetic field and should somehow manifest itself at different latitudes. In particular, the magnetic--field reversal times are expected to be specific in the northern and southern solar hemispheres.

Thus, there are quite a lot of factors that can lead to NSA  of the solar magnetic field. There is no need to change significantly the main scheme of the magnetic field generation. NSA can be caused by the asymmetry in the occurrence of various processes that participate in the dynamo mechanisms. In particular, this can be the asymmetry in the meridional circulation rate \citep[e.g.][]{SC12,BD13,Setal15}  or in the $\alpha$-effect \citep[e.g.][]{Betal13}, the diffusion through the equator at large depths or in the surface layer \citep{Netal14}, or various stochastic effects in the Babcock-Leighton model \citep{Hetal94, GC09, Detal06, Detal07, OK13, Oetal13}. With an appropriate choice of parameters, one will be able to explain both the synchronization of cycles in the two hemispheres and their divergence by up to two years in phase and up to 40 percent in amplitude \citep{Netal14}. However, the question of the origin of small fluctuations in the asymmetry sign remains open. What exactly causes the asymmetry of the components of the generation mechanism is not clear, and we hope that the present study may shed some light on this issue.

Certain differences in the characteristics of differential rotation  in the northern and southern hemispheres have been reported by many authors \citep[e.g.][]{BS06, Zetal11, RUetal12, Letal13}.
Simplified dynamo models consider the arising toroidal field as anti-symmetric in respect to the solar equator while the observational data tell us that the anti-symmetry is far to be perfect.
Later on, the $\alpha$-effect converts this not entirely symmetric toroidal field into the new not entirely symmetric poloidal field. We note again that since the parameter $\alpha$ is rather small, the fluctuations may reach 20 percent of the mean value \citep{SSetal12, Petal12}. Thus, the dependence of the $\alpha$-effect on possible fluctuations and characteristics of the asymmetry may be rather strong. The violation of synchronism in the work of the hemispheres must affect the generation of the poloidal field. Thus, the asymmetry is not only the measure of imbalance of the hemispheres, but, to a certain extent, an indication of the reduced efficiency of generation of the magnetic field of the following cycle.

It is interesting to note that a long--lasting (about 70 yr) imbalance between the hemispheres was observed during the Maunder minimum \citep{RNR93, SNR94, S04}. In fact, only the southern hemisphere was active in that period. Constant prevalence of activity in the southern hemisphere manifested itself in very high values of the asymmetry amplitude. This is consistent with the negative correlation between the asymmetry amplitude and the level of sunspot activity \cite{BO17}.

\begin{figure}
\includegraphics[width=1\columnwidth]{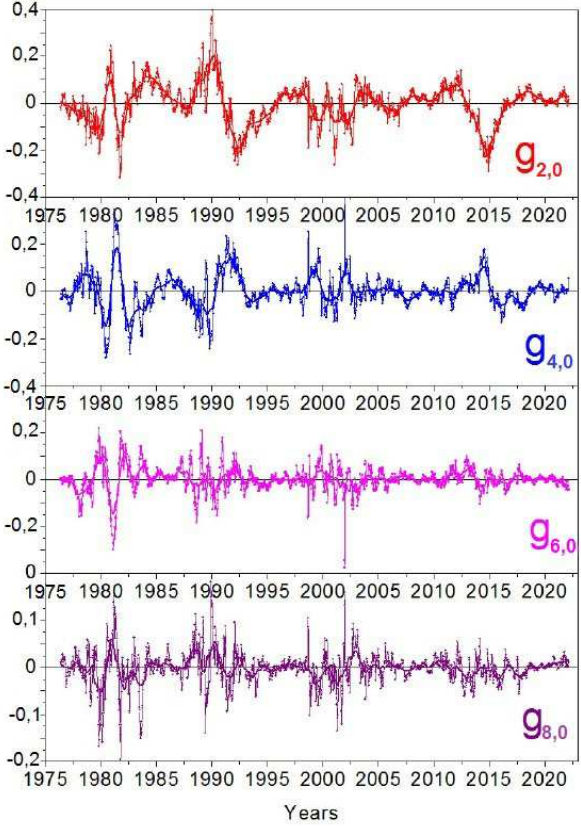}
\caption{Time behaviour of the first four even zonal harmonics. The scales are different for each panel and quite a bit smaller than lower panel of Fig.~1.}
    \label{f4}
\end{figure}

While the above--discussed phenomenon of the  overlapping of cycles is associated with the behaviour of odd zonal harmonics, the asymmetry of the large--scale magnetic field  is determined by even zonal harmonics (the behaviour of the first 4 even zonal harmonics is shown in Fig.~\ref{f4}).
It is interesting that the amplitude of some of the even modes shown in Fig.~4 are more affected by the solar cycle than others, and that this varies from cycle to cycle. E.g. $g_{2,0}$ is quiet in the year 2000, whereas $g_{6,0}$ is quiet in the year 2014.
Comparing Fig.~\ref{f4} with Fig.~\ref{f1}, we learn that the even modes are more variable and, perhaps, more noisy than the odd ones. The main 11-year cycle is quite poorly represented in even modes. Perhaps, some periodicity can be recognized at the middle frequencies; however a spectral analyses is required to verify the expectation (see below). In any case, each of the even harmonics behaves quite specifically. The amplitudes of the even harmonics decay with $l$, but  not as dramatically as the odd ones.

\begin{figure}
\includegraphics[width=1\columnwidth]{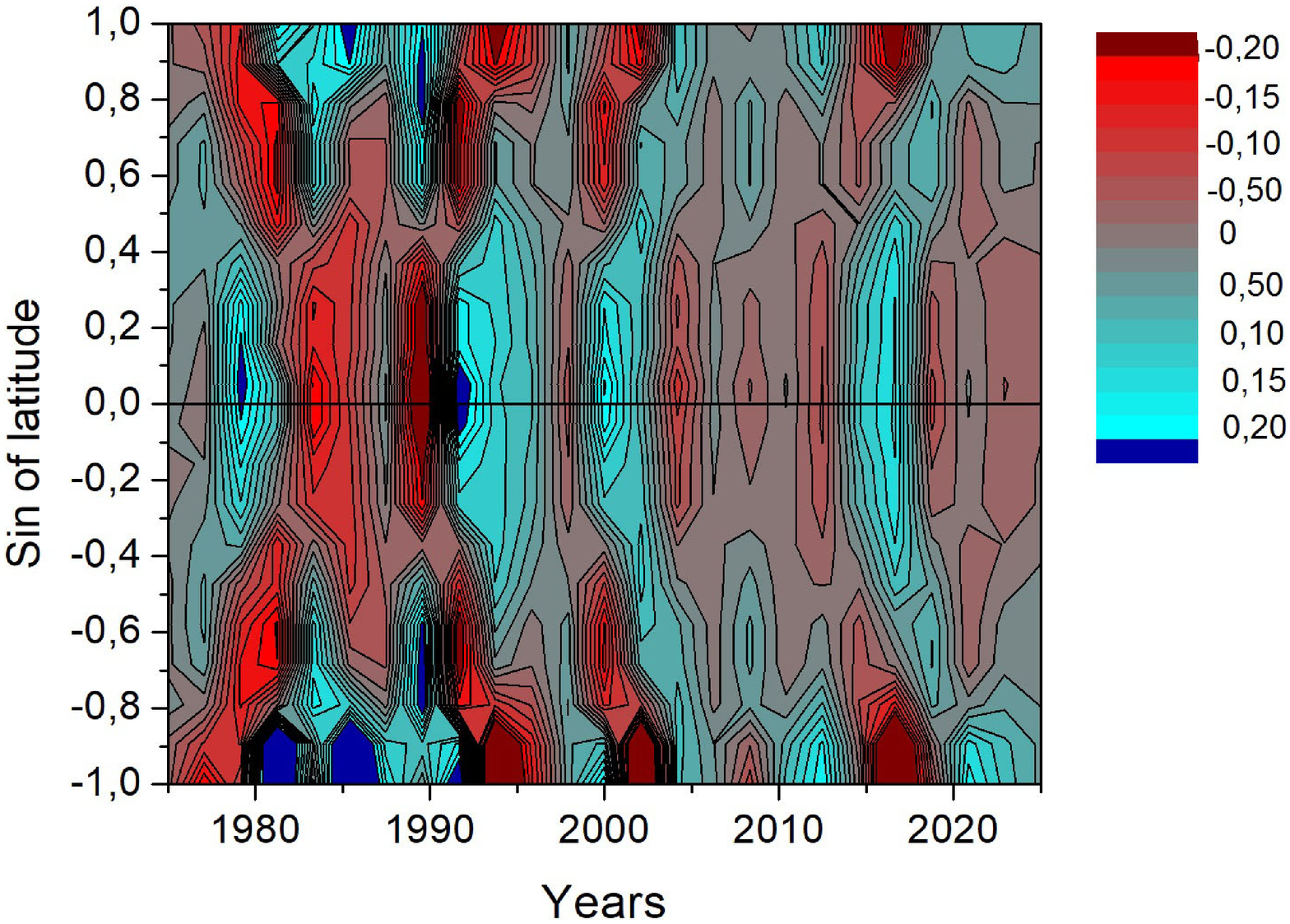}
\includegraphics[width=1\columnwidth]{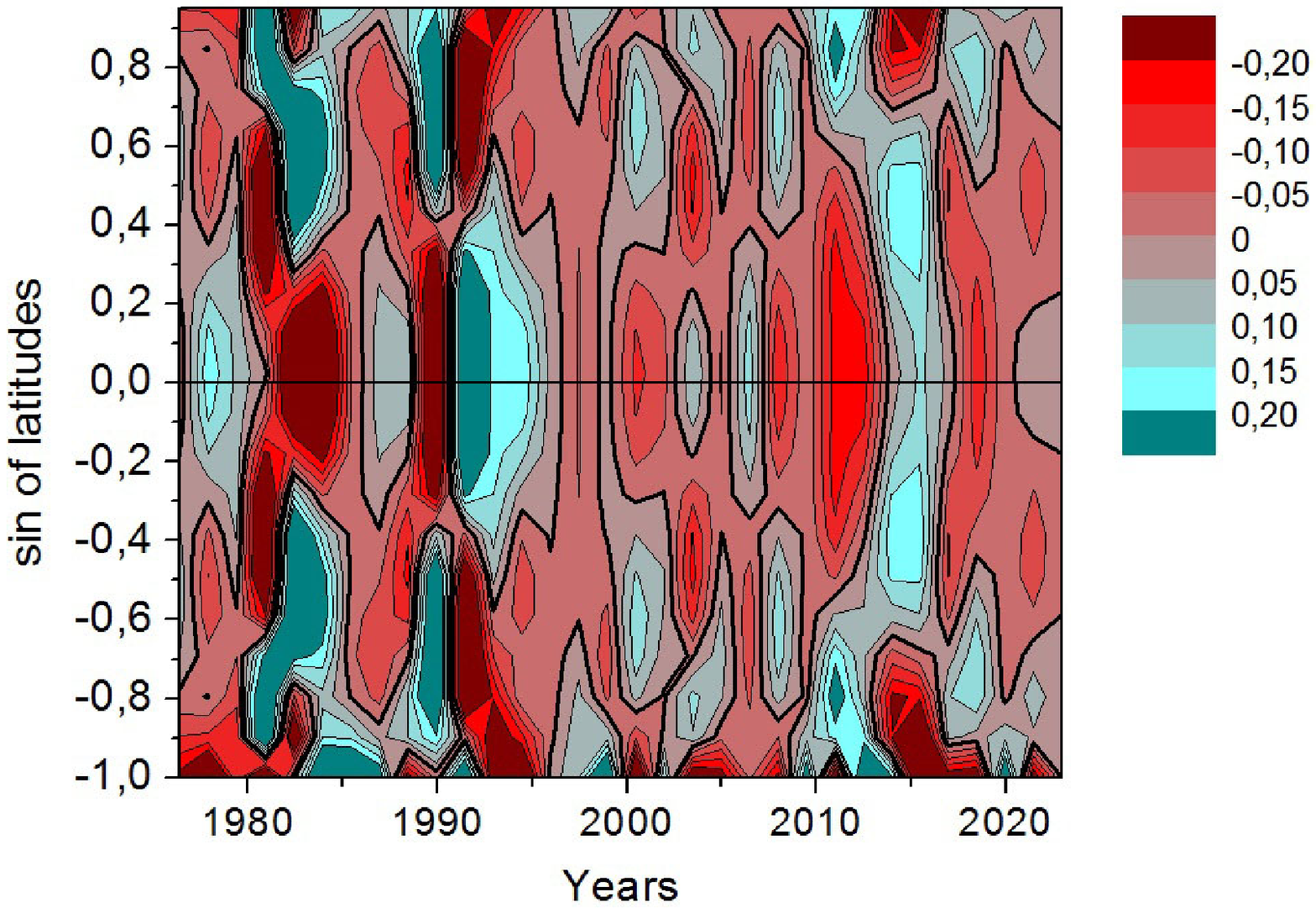}
\caption{Spersynoptic map of the radial magnetic field component produced from two basic even harmonics with $l=2$ and $l=4$. (upper panel) and with four even harmonics (2,4,6,8) - lower panel}
    \label{f5}
\end{figure}

Fig.~\ref{f5} shows a supersynoptic map of the radial magnetic field component produced from two basic even harmonics, $l=2$ and $l=4$, which looks quite different from the synoptic map presented in Fig.~\ref{f3}. Strips about 2--3 years wide are visible, but the polarward drift is not very clearly pronounced. If, however, the drift is visible, the strip has the same width at all latitudes, and the drift also lasts as long as 2--3 years, while traces of longer periodicities are absent. As a result, the sequence of strips with positive and negative polarity has a 5-year periodicity, which holds at all latitudes.

The structure of the bands in Fig. 5 (upper panel) generally resembles the Ripples structure, bands of the same sign drift rapidly towards middle latitudes, and then the drift slows down and the band shifts to a later time. The whole process takes 2-3 years. However, the Ripples in Figure 2 are much thinner. An attempt to improve the situation by including higher harmonics (see Fig. 5, lower panel)  did not improve, but rather worsened the situation. The picture became even more confusing. We have to admit that it is not possible to completely describe the Ripples phenomenon based only on zonal harmonics. This, however, is understandable, since local fields, which are not described by zonal harmonics, apparently play a significant role in the phenomenon of asymmetry. However, the inclusion of harmonics with $m\neq0$ in the analysis is a separate and not simple task, to which we intend to devote a separate article.

Here, we investigate cyclic variations of even harmonics as an asymmetry measure for the large-scale magnetic field. We consider both the amplitudes of harmonics and their absolute values.

\begin{figure}
\begin{tabular}{cc}
\includegraphics[width=40mm]{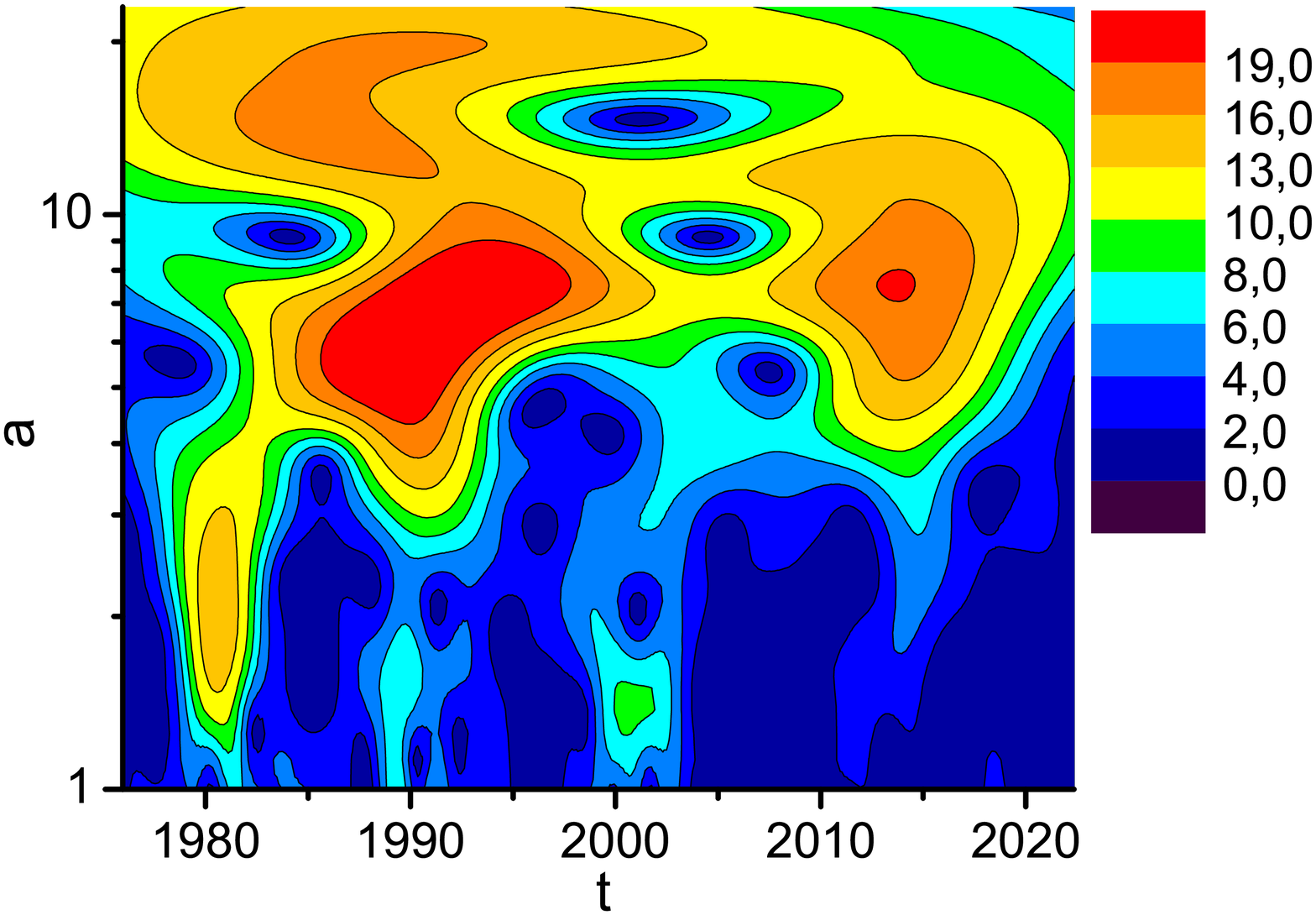} & \includegraphics[width=40mm]{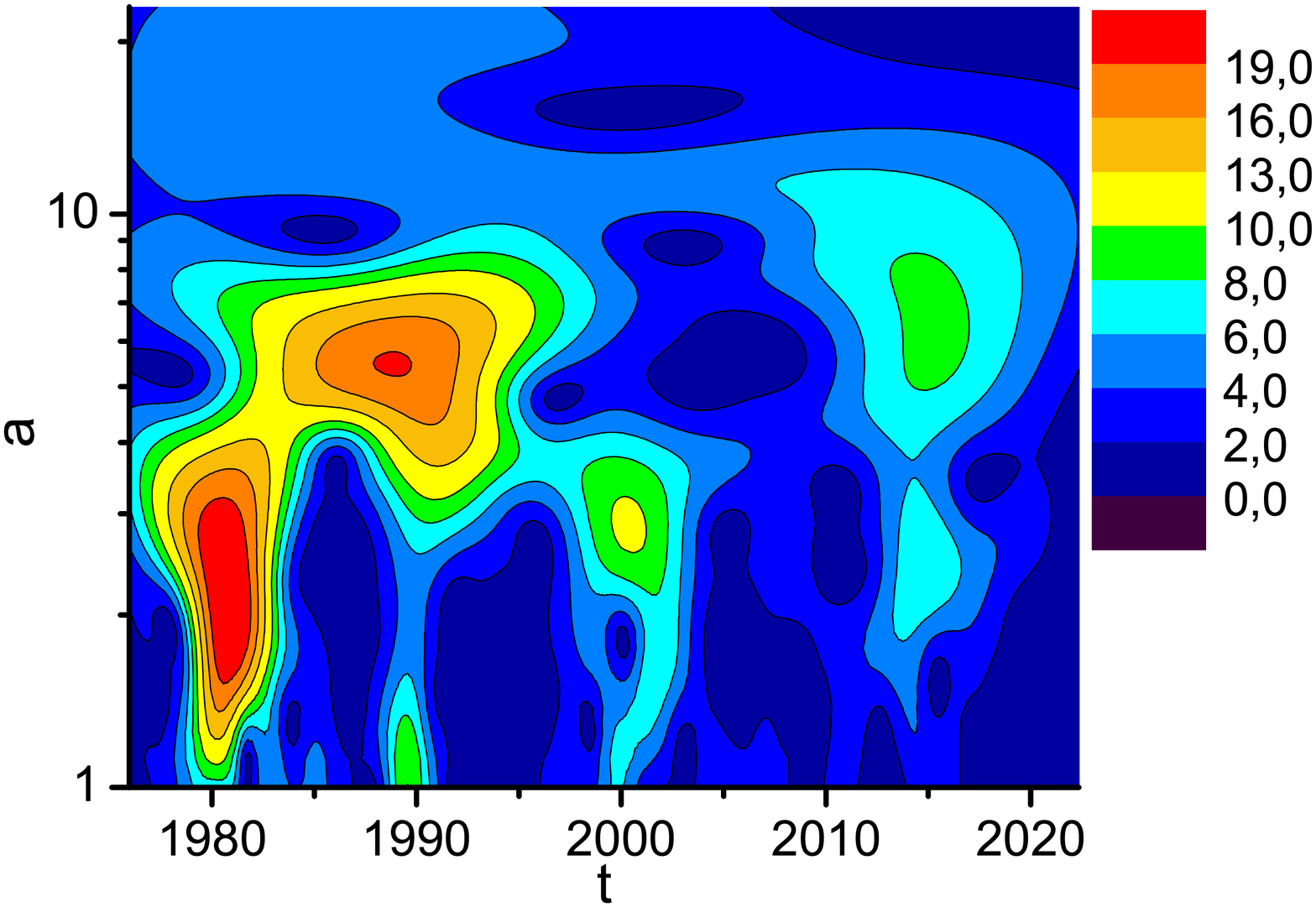}\\
\includegraphics[width=40mm]{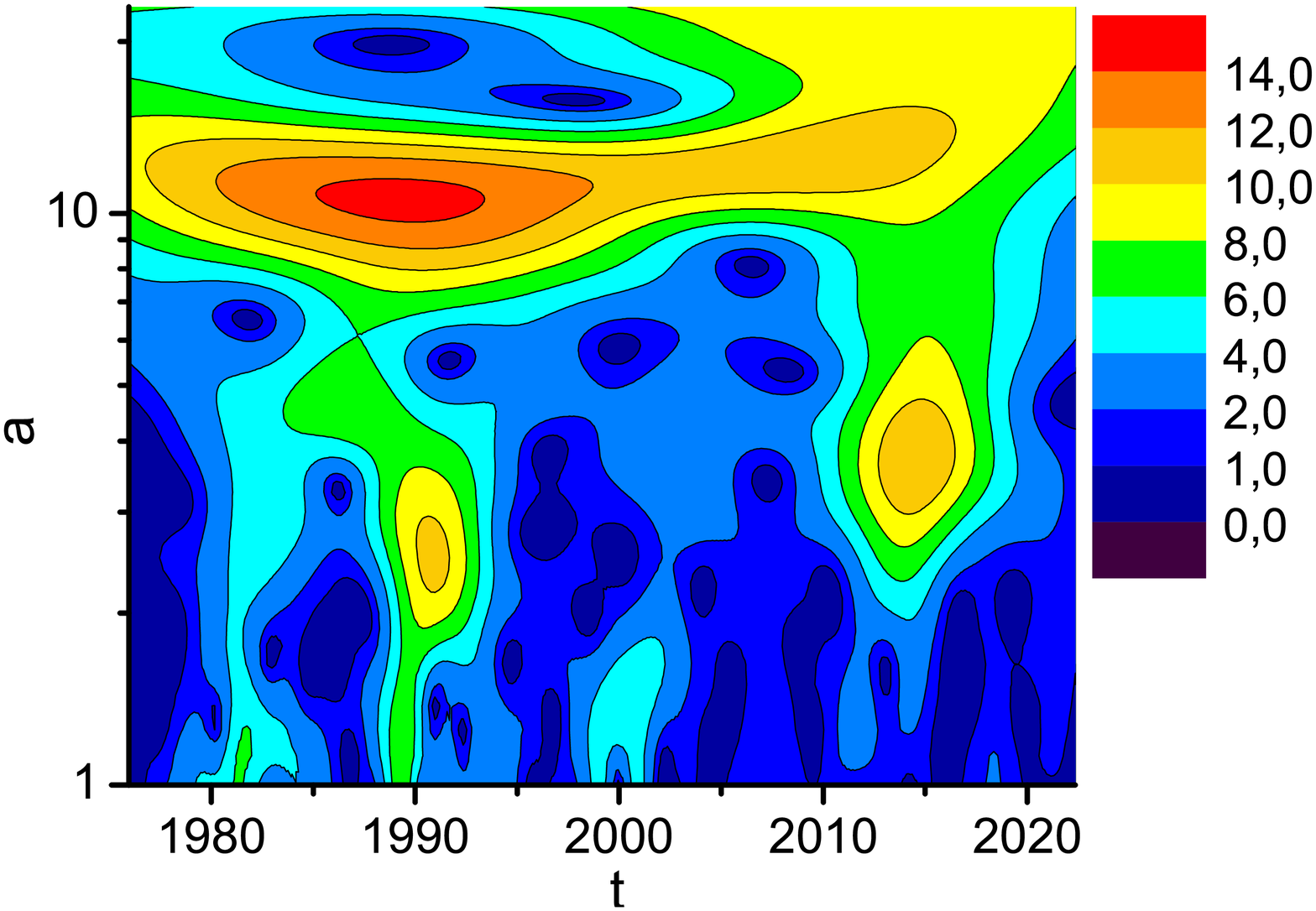} & \includegraphics[width=40mm]{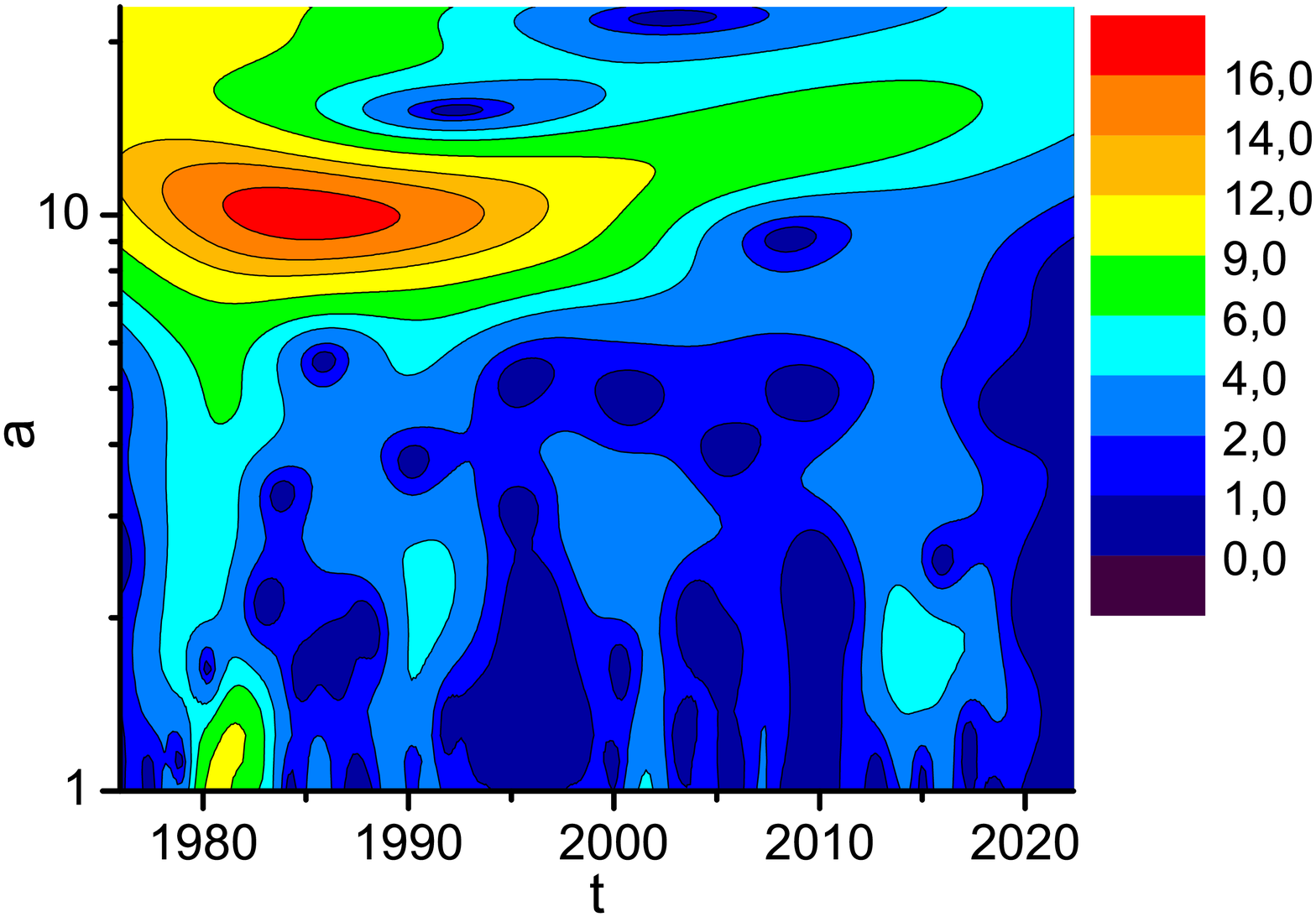}
\end{tabular}
\caption{Wavelet planes for different zonal harmonics, Morlet wavelet is exploited: upper row -- $g_{20}$ (left) and $g_{40}$ (right), lower row -- $|g_{20}|$ (left) and $|g_{40}|$ (right).}
    \label{f6}
\end{figure}

\begin{figure}
\includegraphics[width=1\columnwidth]{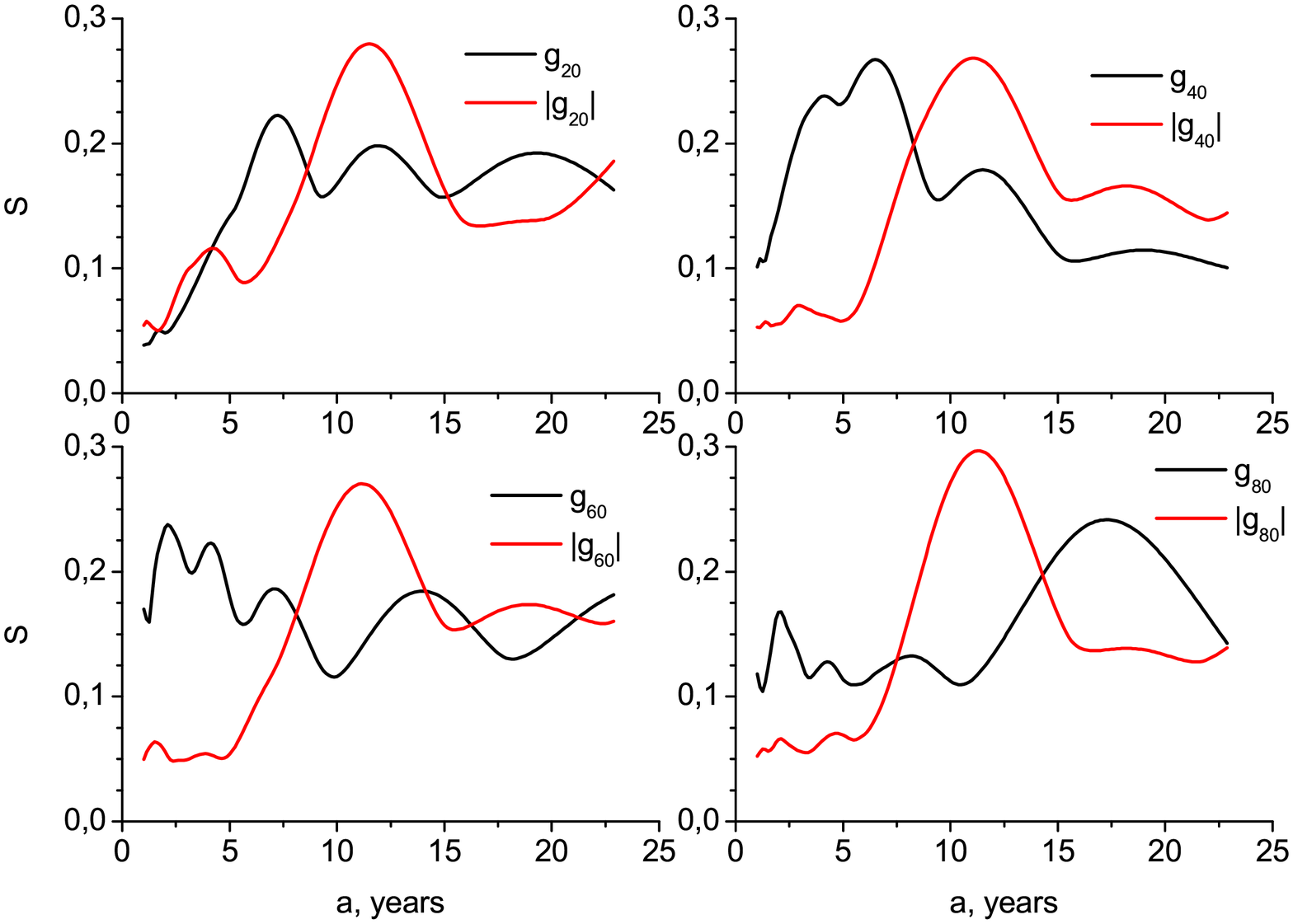}
\caption{Integral wavelet spectra for harmonics $g_{20}$ (top,left) and $g_{40}$ (top, right); $g_{60}$ (bottom, left) and $g_{80}$ (bottom, right) black stands for amplitude, red stands for the absolute value.}
    \label{f7}
\end{figure}

In Fig.~\ref{f6}, we show wavelet planes for different zonal harmonics, Morlet wavelet is exploited: the upper row represents $g_{20}$ (left) and $g_{40}$ right, the lower row shows $|g_{20}|$ (left) and $|g_{40}|$ (right). As expected, the wavelet planes for the harmonic amplitudes show distinct 5--7 year periodicities, while on the plots for absolute values, these periodicities vanish and only the period of about 11 years remains. This is seen even more clearly in Fig.~\ref{f7} with wavelet spectra. The harmonic spectra show mostly a 5 to 6-year periodicity and only hint at an 11-year periodicity, while the absolute value spectra show mostly an 11-year periodicity and only hint at a quasi--biennial oscillation.

\begin{figure}
\includegraphics[width=1\columnwidth]{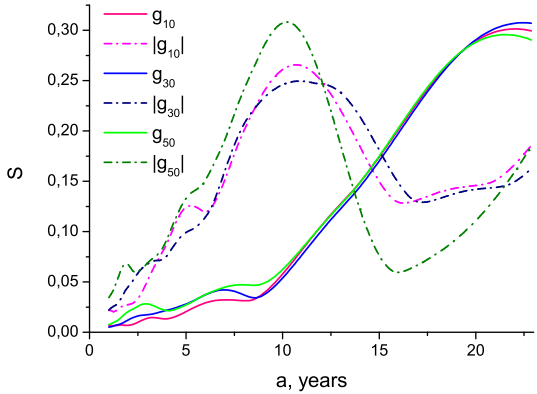}
\caption{Integral wavelet spectra for several odd zonal harmonics and their absolute values.}
    \label{f8}
\end{figure}

It is instructive to compare Fig.~\ref{f7} with Fig.~\ref{f8}, which shows the wavelet spectra of several odd harmonics (to save space, we do not give the corresponding wavelet planes, however they look as expected). Here, the 11-year cycle dominates the spectra of absolute values as it does for the sunspot data, since both tracers do not depend on the polarity of the magnetic field. The signed data hint at a 22-year cycle; however, a reliable analysis is impossible now, as the total length of the available record is 45 years only.

\section{Conclusion and Discussion}

Summarizing the results obtained, we conclude that the basic dynamo paradigm reproduces the main 11-year activity cycle reasonably well in terms of two first odd zonal harmonics of the large-scale magnetic field. On the other hand, the concept of extended cycle needs some clarification as concerns the processes under discussion.  During the  overlapping phase, three activity waves coexist on the solar surface. In this phase lasting about a year (\cite{MIetal14, MIetal19, MIetal20, MIetal21} call it the cycle terminator), along with the polar magnetic field of the previous cycle, a relatively short polar branch (described as "rush to the poles") and a longer equatorward branch appear, the latter existing until the sunspot minimum. As a result, each solar hemisphere contains three activity waves with opposite polarities, which lead to the appearance and enhancement of the odd zonal harmonic with $l=5$. The maximum amplitude of this harmonic dramatically  decayed over the past four cycles similar to the cycle amplitude recorded in sunspot numbers. A particularly strong decline in the value of the $g_{50}$ harmonic is observed after 2000, and this led to a low 24th cycle. Of course, four cycles are not enough to provide a convincing statistics; however it seems plausible that Cycle 25 will not be higher than Cycle 24. Here, we disagree with \cite{MIetal14, MIetal19, MIetal20, MIetal21} who expect a very high Cycle 25. In any case, the overlapping phase took place at the beginning of the 2021, but the amplitude $g_{50}$ remains more or less the same as in Cycle 24.

The SOHO (1996-2010) and SDO (2010-2020) observation data \citep{KPG21, GK22} show that that the development of a new extended solar cycle begins at the latitude of about 60$^{\circ}$ at the base of the convection zone during the maximum of the previous cycle. Then, the process of magnetic field migration to the Sun's surface is divided into two branches: a fast (1--2 years) migration to the poles in the high--latitude zone and a slow migration to the equator at middle and low latitudes for ~10 years. The subsurface rotational shear layer (leptocline) plays a key role in the formation of the magnetic butterfly diagram. A self--consistent MHD model of the solar dynamo developed in the mean--field theory framework is in good qualitative and quantitative agreement with the helioseismic observations. The model shows that the phenomenon of extended solar cycle is due to the magnetic field quenching of the convective heat flux and modulation of the meridional circulation induced by the heat flux variations. The model explains why the polar field at the solar minimum  predicts the following sunspot maximum and points to new possibilities for predicting solar cycles from helioseismological data \citep[e.g.][]{Petal22, Setal21, Betal23}.

Another  effect in the time variation of the large-scale magnetic field  discussed above is not directly related to the 11-year periodicity, but is associated with short-period structures propagating over all latitudes with a moderate poleward drift. The alternation of the sign of the magnetic field  with a characteristic time of $2 \div 5$ years can be seen both directly on the synoptic map and in the results of the wavelet analysis.

 However, when we neglect the sign of the magnetic field and represent the data as an energy spectrum, the amplitude of the 11-year cycle in the spectrum increases sharply, and the high-frequency component becomes much less pronounced. The results are consistent with those obtained by \cite{BO17}. It seems likely that short-term variations both in local magnetic fields and in large-scale ones are determined by the upper part of the convection zone or by the leptocline.

Another point is the relation between the results obtained in this work and the concepts of the solar dynamo  theory. The comparison depends substantially on the level of the dynamo model under consideration. Of course, our results look quite unexpected within the framework of the original solar dynamo models like  \cite{P55, B61} and \cite{L69}. In the traditional cartoons of the solar cycle, it is represented in terms of the first dipole models of the toroidal and poloidal magnetic fields and  a mid-latitude wave of activity propagating equatorward.
In fact, we have a superposition of active phenomena, some of which propagate equatorwards and the others -- polewards with the zonal mode $l=5$ separating both types of propagation. On the other hand, the fact that the solar cycle is something more than just the propagation of magnetic tracers towards the equator has been known since long ago both from the analysis of observations and theoretically. A careful analysis of oversimplified models like \cite{P55} demonstrates that some tracers of poleward propagation are hidden even in these models \citep[e.g.][]{KS95}. It seems quite reasonable to argue that present-day models of the solar dynamo, which introduce additional physical drivers of the solar cycle (such as meridional circulation and/or the possible role of different layers of the solar convection zone in the dynamo activity), actually describe the poleward propagation, together with a well pronounced propagation towards the equator and thereby reflect the main features of the original dynamo models. Both types of propagation are clearly shown in the recent paper by
\citep[e.g.][see Fig.~5 therein]{Petal22}. Presenting our results, we are modelling them to be comparable with the results of \cite{Petal22}. In our opinion, the above plots demonstrate a reasonable agreement between the observations and theory.

Of course we do not insist that that the models by \cite{Petal22} are the only models compatible with the available observations. A comparison of our results with other dynamo models is desirable and could serve a tool for fitting dynamo models to observations; however, this obviously requires a separate study.

Note also that the important role of the zonal harmonics $l=5$ can be to a certain extent explained by simple qualitative arguments. Indeed, one fifth of the meridional size of the hemisphere is about $20^\circ$, which is reasonably close to the width of solar activity wave in a given phase of the solar cycle.

This simple qualitative argument will perhaps make it possible to find out how general the above results are within the framework of the stellar magnetic activity. Indeed, it is difficult to expect that data on stellar activity more or less similar to those used in our study would be available in the foreseeable future; however, an observational estimate of the width of the stellar activity wave for a particular star looks like a more realistic task.

\section*{Acknowledgments}
VNO and DDS acknowledge the support of the Ministry of Science and Higher Education of the Russian Federation under the grant 075-15-2020-780 (VNO) and 075-15-2022-284 (DDS). DDS thanks support by BASIS fund number 21-1-1-4-1. Useful discussions with V.V.Pipin are acknowledged.We are very grateful to our reviewer for very detailed and kind review, which was very helpful for us.

\section{Data availability statement.}
The main data series used in the analysis were obtained at the John Wilcox Stanford Observatory (WSO). These data began in May 1976 (Carrington Rotation 1641) and have been continued till the present (http://wso.stanford.edu/forms/prsyn.html). The polar field is given after http://wso.stanford.edu/Polar.html. In Fig.~2, we have used part of the  picture from https: //solarscience.msfc.nasa.gov/images/magbfly.jpg.

The sunspot data were taken from https://www.sidc.be /silso/datafiles

\bibliographystyle{mnras}
\bibliography{sample631}

\begin{thebibliography}{}
\makeatletter
\relax
\def\mn@urlcharsother{\let\do\@makeother \do\$\do\&\do\#\do\^\do\_\do\%\do\~}
\def\mn@doi{\begingroup\mn@urlcharsother \@ifnextchar [ {\mn@doi@}
  {\mn@doi@[]}}
\def\mn@doi@[#1]#2{\def\@tempa{#1}\ifx\@tempa\@empty \href
  {http://dx.doi.org/#2} {doi:#2}\else \href {http://dx.doi.org/#2} {#1}\fi
  \endgroup}
\def\mn@eprint#1#2{\mn@eprint@#1:#2::\@nil}
\def\mn@eprint@arXiv#1{\href {http://arxiv.org/abs/#1} {{\tt arXiv:#1}}}
\def\mn@eprint@dblp#1{\href {http://dblp.uni-trier.de/rec/bibtex/#1.xml}
  {dblp:#1}}
\def\mn@eprint@#1:#2:#3:#4\@nil{\def\@tempa {#1}\def\@tempb {#2}\def\@tempc
  {#3}\ifx \@tempc \@empty \let \@tempc \@tempb \let \@tempb \@tempa \fi \ifx
  \@tempb \@empty \def\@tempb {arXiv}\fi \@ifundefined
  {mn@eprint@\@tempb}{\@tempb:\@tempc}{\expandafter \expandafter \csname
  mn@eprint@\@tempb\endcsname \expandafter{\@tempc}}}

\bibitem[\protect\citeauthoryear{{Altrock}}{{Altrock}}{1988}]{A88}
{Altrock} R.~C.,  1988, in {Altrock} R.~C.,  ed., Solar and Stellar Coronal
  Structure and Dynamics. pp 414--420

\bibitem[\protect\citeauthoryear{{Altrock}}{{Altrock}}{1997}]{A97}
{Altrock} R.~C.,  1997, \mn@doi [\solphys] {10.1023/A:1004958900477}, \href
  {https://ui.adsabs.harvard.edu/abs/1997SoPh..170..411A} {170, 411}

\bibitem[\protect\citeauthoryear{{Babcock}}{{Babcock}}{1961}]{B61}
{Babcock} H.~W.,  1961, \mn@doi [\apj] {10.1086/147060}, \href
  {https://ui.adsabs.harvard.edu/abs/1961ApJ...133..572B} {133, 572}

\bibitem[\protect\citeauthoryear{{Badalyan}}{{Badalyan}}{2011}]{B11}
{Badalyan} O.~G.,  2011, \mn@doi [Astronomy Reports]
  {10.1134/S1063772911100027}, \href
  {https://ui.adsabs.harvard.edu/abs/2011ARep...55..928B} {55, 928}

\bibitem[\protect\citeauthoryear{{Badalyan} \& {Obridko}}{{Badalyan} \&
  {Obridko}}{2011}]{BO11}
{Badalyan} O.~G.,  {Obridko} V.~N.,  2011, \mn@doi [\na]
  {10.1016/j.newast.2011.01.005}, \href
  {https://ui.adsabs.harvard.edu/abs/2011NewA...16..357B} {16, 357}

\bibitem[\protect\citeauthoryear{{Badalyan} \& {Obridko}}{{Badalyan} \&
  {Obridko}}{2017}]{BO17}
{Badalyan} O.~G.,  {Obridko} V.~N.,  2017, \mn@doi [\aap]
  {10.1051/0004-6361/201527790}, \href
  {https://ui.adsabs.harvard.edu/abs/2017A&A...603A.109B} {603, A109}

\bibitem[\protect\citeauthoryear{{Badalyan} \& {S{\'y}kora}}{{Badalyan} \&
  {S{\'y}kora}}{2006}]{BS06}
{Badalyan} O.~G.,  {S{\'y}kora} J.,  2006, \mn@doi [Advances in Space Research]
  {10.1016/j.asr.2006.03.026}, \href
  {https://ui.adsabs.harvard.edu/abs/2006AdSpR..38..906B} {38, 906}

\bibitem[\protect\citeauthoryear{{Badalyan}, {Obridko}, {Ryb{\'a}k}  \&
  {S{\'y}kora}}{{Badalyan} et~al.}{2005}]{Betal05}
{Badalyan} O.~G.,  {Obridko} V.~N.,  {Ryb{\'a}k} J.,   {S{\'y}kora} J.,  2005,
  \mn@doi [Astronomy Reports] {10.1134/1.2010655}, \href
  {https://ui.adsabs.harvard.edu/abs/2005ARep...49..659B} {49, 659}

\bibitem[\protect\citeauthoryear{{Badalyan}, {Obridko}  \&
  {S{\'y}kora}}{{Badalyan} et~al.}{2008}]{Betal08}
{Badalyan} O.~G.,  {Obridko} V.~N.,   {S{\'y}kora} J.,  2008, \mn@doi
  [\solphys] {10.1007/s11207-008-9120-0}, \href
  {https://ui.adsabs.harvard.edu/abs/2008SoPh..247..379B} {247, 379}

\bibitem[\protect\citeauthoryear{{Belucz} \& {Dikpati}}{{Belucz} \&
  {Dikpati}}{2013}]{BD13}
{Belucz} B.,  {Dikpati} M.,  2013, \mn@doi [\apj] {10.1088/0004-637X/779/1/4},
  \href {https://ui.adsabs.harvard.edu/abs/2013ApJ...779....4B} {779, 4}

\bibitem[\protect\citeauthoryear{{Belucz }, {Forg{\'a}cs-Dajka}  \&
  {Dikpati}}{{Belucz } et~al.}{2013}]{Betal13}
{Belucz } B.,  {Forg{\'a}cs-Dajka} E.,   {Dikpati} M.,  2013, \mn@doi
  [Astronomische Nachrichten] {10.1002/asna.201211970}, \href
  {https://ui.adsabs.harvard.edu/abs/2013AN....334..960B} {334, 960}

\bibitem[\protect\citeauthoryear{{Bocchino}}{{Bocchino}}{1933}]{B33}
{Bocchino} G.,  1933, \memsai, \href
  {https://ui.adsabs.harvard.edu/abs/1933MmSAI...6..479B} {6, 479}

\bibitem[\protect\citeauthoryear{{Boyer} \& {Levy}}{{Boyer} \&
  {Levy}}{1984}]{BL84}
{Boyer} D.~W.,  {Levy} E.~H.,  1984, \mn@doi [\apj] {10.1086/161755}, \href
  {https://ui.adsabs.harvard.edu/abs/1984ApJ...277..848B} {277, 848}

\bibitem[\protect\citeauthoryear{{Brandenburg}, {Elstner}, {Masada}  \&
  {Pipin}}{{Brandenburg} et~al.}{2023}]{Betal23}
{Brandenburg} A.,  {Elstner} D.,  {Masada} Y.,   {Pipin} V.,  2023, \mn@doi
  [arXiv e-prints] {10.48550/arXiv.2303.12425}, \href
  {https://ui.adsabs.harvard.edu/abs/2023arXiv230312425B} {p. arXiv:2303.12425}

\bibitem[\protect\citeauthoryear{{Cameron} \& {Sch{\"u}ssler}}{{Cameron} \&
  {Sch{\"u}ssler}}{2012}]{SC12}
{Cameron} R.~H.,  {Sch{\"u}ssler} M.,  2012, \mn@doi [\aap]
  {10.1051/0004-6361/201219914}, \href
  {https://ui.adsabs.harvard.edu/abs/2012A&A...548A..57C} {548, A57}

\bibitem[\protect\citeauthoryear{{Carbonell}, {Oliver}  \&
  {Ballester}}{{Carbonell} et~al.}{1993}]{Cetal93}
{Carbonell} M.,  {Oliver} R.,   {Ballester} J.~L.,  1993, \aap, \href
  {https://ui.adsabs.harvard.edu/abs/1993A&A...274..497C} {274, 497}

\bibitem[\protect\citeauthoryear{{Carbonell}, {Terradas}, {Oliver}  \&
  {Ballester}}{{Carbonell} et~al.}{2007}]{Cetal07}
{Carbonell} M.,  {Terradas} J.,  {Oliver} R.,   {Ballester} J.~L.,  2007,
  \mn@doi [\aap] {10.1051/0004-6361:20078004}, \href
  {https://ui.adsabs.harvard.edu/abs/2007A&A...476..951C} {476, 951}

\bibitem[\protect\citeauthoryear{{Dikpati}, {de Toma}  \& {Gilman}}{{Dikpati}
  et~al.}{2006}]{Detal06}
{Dikpati} M.,  {de Toma} G.,   {Gilman} P.~A.,  2006, \mn@doi [\grl]
  {10.1029/2005GL025221}, \href
  {https://ui.adsabs.harvard.edu/abs/2006GeoRL..33.5102D} {33, L05102}

\bibitem[\protect\citeauthoryear{{Dikpati}, {Gilman}, {de Toma}  \&
  {Ghosh}}{{Dikpati} et~al.}{2007}]{Detal07}
{Dikpati} M.,  {Gilman} P.~A.,  {de Toma} G.,   {Ghosh} S.~S.,  2007, \mn@doi
  [\solphys] {10.1007/s11207-007-9016-4}, \href
  {https://ui.adsabs.harvard.edu/abs/2007SoPh..245....1D} {245, 1}

\bibitem[\protect\citeauthoryear{{Getling} \& {Kosovichev}}{{Getling} \&
  {Kosovichev}}{2022}]{GK22}
{Getling} A.~V.,  {Kosovichev} A.~G.,  2022, \mn@doi [\apj]
  {10.3847/1538-4357/ac8870}, \href
  {https://ui.adsabs.harvard.edu/abs/2022ApJ...937...41G} {937, 41}

\bibitem[\protect\citeauthoryear{{Goel} \& {Choudhuri}}{{Goel} \&
  {Choudhuri}}{2009}]{GC09}
{Goel} A.,  {Choudhuri} A.~R.,  2009, \mn@doi [Research in Astronomy and
  Astrophysics] {10.1088/1674-4527/9/1/010}, \href
  {https://ui.adsabs.harvard.edu/abs/2009RAA.....9..115G} {9, 115}

\bibitem[\protect\citeauthoryear{{Hansen} \& {Hansen}}{{Hansen} \&
  {Hansen}}{1975}]{HH75}
{Hansen} R.,  {Hansen} S.,  1975, \mn@doi [\solphys] {10.1007/BF00156857},
  \href {https://ui.adsabs.harvard.edu/abs/1975SoPh...44..225H} {44, 225}

\bibitem[\protect\citeauthoryear{{Harvey} \& {Martin}}{{Harvey} \&
  {Martin}}{1973}]{HM73}
{Harvey} K.~L.,  {Martin} S.~F.,  1973, \mn@doi [\solphys]
  {10.1007/BF00154951}, \href
  {https://ui.adsabs.harvard.edu/abs/1973SoPh...32..389H} {32, 389}

\bibitem[\protect\citeauthoryear{{Hathaway} \& {Wilson}}{{Hathaway} \&
  {Wilson}}{1990}]{HW90}
{Hathaway} D.~H.,  {Wilson} R.~M.,  1990, \mn@doi [\apj] {10.1086/168913},
  \href {https://ui.adsabs.harvard.edu/abs/1990ApJ...357..271H} {357, 271}

\bibitem[\protect\citeauthoryear{{Hoeksema} \& {Scherrer}}{{Hoeksema} \&
  {Scherrer}}{1986}]{HS86}
{Hoeksema} J.~T.,  {Scherrer} P.~H.,  1986, WDCA Report UAG-94, NGDC, Boulder,
  CO

\bibitem[\protect\citeauthoryear{{Howard} \& {Labonte}}{{Howard} \&
  {Labonte}}{1980}]{HL80}
{Howard} R.,  {Labonte} B.~J.,  1980, \mn@doi [\apjl] {10.1086/183286}, \href
  {https://ui.adsabs.harvard.edu/abs/1980ApJ...239L..33H} {239, L33}

\bibitem[\protect\citeauthoryear{{Hoyng}, {Schmitt}  \& {Teuben}}{{Hoyng}
  et~al.}{1994}]{Hetal94}
{Hoyng} P.,  {Schmitt} D.,   {Teuben} L.~J.~W.,  1994, \aap, \href
  {https://ui.adsabs.harvard.edu/abs/1994A&A...289..265H} {289, 265}

\bibitem[\protect\citeauthoryear{{Kambry} \& {Nishikawa}}{{Kambry} \&
  {Nishikawa}}{1990}]{KN90}
{Kambry} M.~A.,  {Nishikawa} J.,  1990, \mn@doi [\solphys]
  {10.1007/BF00158300}, \href
  {https://ui.adsabs.harvard.edu/abs/1990SoPh..126...89K} {126, 89}

\bibitem[\protect\citeauthoryear{{Kharshiladze} \& {Ivanov}}{{Kharshiladze} \&
  {Ivanov}}{1994}]{KI94}
{Kharshiladze} A.~F.,  {Ivanov} K.~G.,  1994, Geomagnetism and Aeronomy, \href
  {https://ui.adsabs.harvard.edu/abs/1994Ge&Ae..34...22K} {34, 22}

\bibitem[\protect\citeauthoryear{{Kosovichev}, {Pipin}  \&
  {Getling}}{{Kosovichev} et~al.}{2021}]{KPG21}
{Kosovichev} A.,  {Pipin} V.,   {Getling} A.,  2021, in American Astronomical
  Society Meeting Abstracts. p. 304.05

\bibitem[\protect\citeauthoryear{{Kuzanyan} \& {Sokoloff}}{{Kuzanyan} \&
  {Sokoloff}}{1995}]{KS95}
{Kuzanyan} K.~M.,  {Sokoloff} D.~D.,  1995, \mn@doi [Geophysical and
  Astrophysical Fluid Dynamics] {10.1080/03091929508229073}, \href
  {https://ui.adsabs.harvard.edu/abs/1995GApFD..81..113K} {81, 113}

\bibitem[\protect\citeauthoryear{{Legrand} \& {Simon}}{{Legrand} \&
  {Simon}}{1981}]{LS81}
{Legrand} J.~P.,  {Simon} P.~A.,  1981, \mn@doi [\solphys]
  {10.1007/BF00154399}, \href
  {https://ui.adsabs.harvard.edu/abs/1981SoPh...70..173L} {70, 173}

\bibitem[\protect\citeauthoryear{{Leighton}}{{Leighton}}{1969}]{L69}
{Leighton} R.~B.,  1969, \mn@doi [\apj] {10.1086/149943}, \href
  {https://ui.adsabs.harvard.edu/abs/1969ApJ...156....1L} {156, 1}

\bibitem[\protect\citeauthoryear{{Leroy} \& {Noens}}{{Leroy} \&
  {Noens}}{1983}]{LN83}
{Leroy} J.~L.,  {Noens} J.~C.,  1983, \aap, \href
  {https://ui.adsabs.harvard.edu/abs/1983A&A...120L...1L} {120, L1}

\bibitem[\protect\citeauthoryear{{Li}, {Wang}, {Xiong}, {Liang}, {Yun}  \&
  {Gu}}{{Li} et~al.}{2002}]{Letal02}
{Li} K.~J.,  {Wang} J.~X.,  {Xiong} S.~Y.,  {Liang} H.~F.,  {Yun} H.~S.,   {Gu}
  X.~M.,  2002, \mn@doi [\aap] {10.1051/0004-6361:20011799}, \href
  {https://ui.adsabs.harvard.edu/abs/2002A&A...383..648L} {383, 648}

\bibitem[\protect\citeauthoryear{{Li}, {Shi}, {Xie}, {Gao}, {Liang}, {Zhan}  \&
  {Feng}}{{Li} et~al.}{2013}]{Letal13}
{Li} K.~J.,  {Shi} X.~J.,  {Xie} J.~L.,  {Gao} P.~X.,  {Liang} H.~F.,  {Zhan}
  L.~S.,   {Feng} W.,  2013, \mn@doi [\mnras] {10.1093/mnras/stt744}, \href
  {https://ui.adsabs.harvard.edu/abs/2013MNRAS.433..521L} {433, 521}

\bibitem[\protect\citeauthoryear{{Maris}, {Popescu}  \& {Mierla}}{{Maris}
  et~al.}{2002}]{Metal02}
{Maris} G.,  {Popescu} M.~D.,   {Mierla} M.,  2002, Romanian Astronomical
  Journal, \href {https://ui.adsabs.harvard.edu/abs/2002RoAJ...12..131M} {12,
  131}

\bibitem[\protect\citeauthoryear{{Mayaud}}{{Mayaud}}{1975}]{M75}
{Mayaud} P.~N.,  1975, \mn@doi [\jgr] {10.1029/JA080i001p00111}, \href
  {https://ui.adsabs.harvard.edu/abs/1975JGR....80..111M} {80, 111}

\bibitem[\protect\citeauthoryear{{McIntosh}}{{McIntosh}}{1992}]{MI92}
{McIntosh} P.~S.,  1992, in {Harvey} K.~L.,  ed.,  Astronomical Society of the
  Pacific Conference Series Vol. 27, The Solar Cycle. p.~14

\bibitem[\protect\citeauthoryear{{McIntosh} et~al.,}{{McIntosh}
  et~al.}{2014}]{MIetal14}
{McIntosh} S.~W.,  et~al., 2014, \mn@doi [\apj] {10.1088/0004-637X/792/1/12},
  \href {https://ui.adsabs.harvard.edu/abs/2014ApJ...792...12M} {792, 12}

\bibitem[\protect\citeauthoryear{{McIntosh}, {Leamon}, {Egeland}, {Dikpati},
  {Fan}  \& {Rempel}}{{McIntosh} et~al.}{2019}]{MIetal19}
{McIntosh} S.~W.,  {Leamon} R.~J.,  {Egeland} R.,  {Dikpati} M.,  {Fan} Y.,
  {Rempel} M.,  2019, \mn@doi [\solphys] {10.1007/s11207-019-1474-y}, \href
  {https://ui.adsabs.harvard.edu/abs/2019SoPh..294...88M} {294, 88}

\bibitem[\protect\citeauthoryear{{McIntosh}, {Chapman}, {Leamon}, {Egeland}  \&
  {Watkins}}{{McIntosh} et~al.}{2020}]{MIetal20}
{McIntosh} S.~W.,  {Chapman} S.,  {Leamon} R.~J.,  {Egeland} R.,   {Watkins}
  N.~W.,  2020, \mn@doi [\solphys] {10.1007/s11207-020-01723-y}, \href
  {https://ui.adsabs.harvard.edu/abs/2020SoPh..295..163M} {295, 163}

\bibitem[\protect\citeauthoryear{{McIntosh} et~al.,}{{McIntosh}
  et~al.}{2021}]{MIetal21}
{McIntosh} S.~W.,  et~al., 2021, \mn@doi [\solphys]
  {10.1007/s11207-021-01938-7}, \href
  {https://ui.adsabs.harvard.edu/abs/2021SoPh..296..189M} {296, 189}

\bibitem[\protect\citeauthoryear{{Mordvinov}}{{Mordvinov}}{2007}]{M07}
{Mordvinov} A.~V.,  2007, \mn@doi [\solphys] {10.1007/s11207-007-9082-7}, \href
  {https://ui.adsabs.harvard.edu/abs/2007SoPh..246..445M} {246, 445}

\bibitem[\protect\citeauthoryear{{Norton}, {Charbonneau}  \& {Passos}}{{Norton}
  et~al.}{2014}]{Netal14}
{Norton} A.~A.,  {Charbonneau} P.,   {Passos} D.,  2014, \mn@doi [\ssr]
  {10.1007/s11214-014-0100-4}, \href
  {https://ui.adsabs.harvard.edu/abs/2014SSRv..186..251N} {186, 251}

\bibitem[\protect\citeauthoryear{{Obridko} \& {Shelting}}{{Obridko} \&
  {Shelting}}{1992}]{OS92}
{Obridko} V.~N.,  {Shelting} B.~D.,  1992, \mn@doi [\solphys]
  {10.1007/BF00146582}, \href
  {https://ui.adsabs.harvard.edu/abs/1992SoPh..137..167O} {137, 167}

\bibitem[\protect\citeauthoryear{{Obridko} \& {Shelting}}{{Obridko} \&
  {Shelting}}{2000a}]{OS00b}
{Obridko} V.~N.,  {Shelting} B.~D.,  2000a, \mn@doi [Astronomy Reports]
  {10.1134/1.163828}, \href
  {https://ui.adsabs.harvard.edu/abs/2000ARep...44..103O} {44, 103}

\bibitem[\protect\citeauthoryear{{Obridko} \& {Shelting}}{{Obridko} \&
  {Shelting}}{2000b}]{OS00a}
{Obridko} V.~N.,  {Shelting} B.~D.,  2000b, \mn@doi [Astronomy Reports]
  {10.1134/1.163849}, \href
  {https://ui.adsabs.harvard.edu/abs/2000ARep...44..262O} {44, 262}

\bibitem[\protect\citeauthoryear{{Obridko} \& {Shelting}}{{Obridko} \&
  {Shelting}}{2016}]{OS16}
{Obridko} V.~N.,  {Shelting} B.~D.,  2016, \mn@doi [Astronomy Letters]
  {10.1134/S1063773716080041}, \href
  {https://ui.adsabs.harvard.edu/abs/2016AstL...42..631O} {42, 631}

\bibitem[\protect\citeauthoryear{{Obridko} \& {Yermakov}}{{Obridko} \&
  {Yermakov}}{1989}]{OY89}
{Obridko} V.~N.,  {Yermakov} F.~A.,  1989, Astronomicheskij Tsirkulyar, \href
  {https://ui.adsabs.harvard.edu/abs/1989ATsir1539...24O} {1539, 24}

\bibitem[\protect\citeauthoryear{{Obridko}, {Sokoloff}, {Shelting}, {Shibalova}
   \& {Livshits}}{{Obridko} et~al.}{2020}]{Betal20}
{Obridko} V.~N.,  {Sokoloff} D.~D.,  {Shelting} B.~D.,  {Shibalova} A.~S.,
  {Livshits} I.~M.,  2020, \mn@doi [\mnras] {10.1093/mnras/staa147}, \href
  {https://ui.adsabs.harvard.edu/abs/2020MNRAS.492.5582O} {492, 5582}

\bibitem[\protect\citeauthoryear{{Obridko}, {Sokoloff}, {Pipin}, {Shibalova}
  \& {Livshits}}{{Obridko} et~al.}{2021}]{Obetal21}
{Obridko} V.~N.,  {Sokoloff} D.~D.,  {Pipin} V.~V.,  {Shibalova} A.~S.,
  {Livshits} I.~M.,  2021, \mn@doi [Journal of Atmospheric and
  Solar–Terrestrial Physics] {10.1016/j.jastp.2021.105743}, 225

\bibitem[\protect\citeauthoryear{{Olemskoy} \& {Kitchatinov}}{{Olemskoy} \&
  {Kitchatinov}}{2013}]{OK13}
{Olemskoy} S.~V.,  {Kitchatinov} L.~L.,  2013, \mn@doi [\apj]
  {10.1088/0004-637X/777/1/71}, \href
  {https://ui.adsabs.harvard.edu/abs/2013ApJ...777...71O} {777, 71}

\bibitem[\protect\citeauthoryear{{Olemskoy}, {Choudhuri}  \&
  {Kitchatinov}}{{Olemskoy} et~al.}{2013}]{Oetal13}
{Olemskoy} S.~V.,  {Choudhuri} A.~R.,   {Kitchatinov} L.~L.,  2013, \mn@doi
  [Astronomy Reports] {10.1134/S1063772913050065}, \href
  {https://ui.adsabs.harvard.edu/abs/2013ARep...57..458O} {57, 458}

\bibitem[\protect\citeauthoryear{{Parker}}{{Parker}}{1955}]{P55}
{Parker} E.~N.,  1955, \mn@doi [\apj] {10.1086/146087}, \href
  {https://ui.adsabs.harvard.edu/abs/1955ApJ...122..293P} {122, 293}

\bibitem[\protect\citeauthoryear{{Pipin}, {Sokoloff}  \& {Usoskin}}{{Pipin}
  et~al.}{2012}]{Petal12}
{Pipin} V.~V.,  {Sokoloff} D.~D.,   {Usoskin} I.~G.,  2012, \mn@doi [\aap]
  {10.1051/0004-6361/201118733}, \href
  {https://ui.adsabs.harvard.edu/abs/2012A&A...542A..26P} {542, A26}

\bibitem[\protect\citeauthoryear{{Pipin}, {Kosovichev}  \& {Tomin}}{{Pipin}
  et~al.}{2022}]{Petal22}
{Pipin} V.~V.,  {Kosovichev} A.~G.,   {Tomin} V.~E.,  2022, \mn@doi [arXiv
  e-prints] {10.48550/arXiv.2210.08764}, \href
  {https://ui.adsabs.harvard.edu/abs/2022arXiv221008764P} {p. arXiv:2210.08764}

\bibitem[\protect\citeauthoryear{{Ribes} \& {Nesme-Ribes}}{{Ribes} \&
  {Nesme-Ribes}}{1993}]{RNR93}
{Ribes} J.~C.,  {Nesme-Ribes} E.,  1993, \aap, \href
  {https://ui.adsabs.harvard.edu/abs/1993A&A...276..549R} {276, 549}

\bibitem[\protect\citeauthoryear{{Rightmire-Upton}, {Hathaway}  \&
  {Kosak}}{{Rightmire-Upton} et~al.}{2012}]{RUetal12}
{Rightmire-Upton} L.,  {Hathaway} D.~H.,   {Kosak} K.,  2012, \mn@doi [\apjl]
  {10.1088/2041-8205/761/1/L14}, \href
  {https://ui.adsabs.harvard.edu/abs/2012ApJ...761L..14R} {761, L14}

\bibitem[\protect\citeauthoryear{{Shetye}, {Tripathi}  \& {Dikpati}}{{Shetye}
  et~al.}{2015}]{Setal15}
{Shetye} J.,  {Tripathi} D.,   {Dikpati} M.,  2015, \mn@doi [\apj]
  {10.1088/0004-637X/799/2/220}, \href
  {https://ui.adsabs.harvard.edu/abs/2015ApJ...799..220S} {799, 220}

\bibitem[\protect\citeauthoryear{{Snodgrass} \& {Wilson}}{{Snodgrass} \&
  {Wilson}}{1987}]{SW87}
{Snodgrass} H.~B.,  {Wilson} P.~R.,  1987, \mn@doi [\nat] {10.1038/328696a0},
  \href {https://ui.adsabs.harvard.edu/abs/1987Natur.328..696S} {328, 696}

\bibitem[\protect\citeauthoryear{{Sokoloff}}{{Sokoloff}}{2004}]{S04}
{Sokoloff} D.,  2004, \mn@doi [\solphys] {10.1007/s11207-005-4176-6}, \href
  {https://ui.adsabs.harvard.edu/abs/2004SoPh..224..145S} {224, 145}

\bibitem[\protect\citeauthoryear{{Sokoloff} \& {Nesme-Ribes}}{{Sokoloff} \&
  {Nesme-Ribes}}{1994}]{SNR94}
{Sokoloff} D.,  {Nesme-Ribes} E.,  1994, \aap, \href
  {https://ui.adsabs.harvard.edu/abs/1994A&A...288..293S} {288, 293}

\bibitem[\protect\citeauthoryear{{Sokoloff}, {Sobko}, {Trukhin}  \&
  {Zadkov}}{{Sokoloff} et~al.}{2012}]{SSetal12}
{Sokoloff} D.~D.,  {Sobko} G.~S.,  {Trukhin} V.~I.,   {Zadkov} V.~N.,  2012, in
  {Mandrini} C.~H.,  {Webb} D.~F.,  eds, ~ Vol. 286, Comparative Magnetic
  Minima: Characterizing Quiet Times in the Sun and Stars. pp 360--366,
  \mn@doi{10.1017/S1743921312005091}

\bibitem[\protect\citeauthoryear{{Stejko}, {Kosovichev}  \& {Pipin}}{{Stejko}
  et~al.}{2021}]{Setal21}
{Stejko} A.~M.,  {Kosovichev} A.~G.,   {Pipin} V.~V.,  2021, \mn@doi [\apj]
  {10.3847/1538-4357/abec70}, \href
  {https://ui.adsabs.harvard.edu/abs/2021ApJ...911...90S} {911, 90}

\bibitem[\protect\citeauthoryear{{S{\'y}kora} \& {Ryb{\'a}k}}{{S{\'y}kora} \&
  {Ryb{\'a}k}}{2010}]{SR10}
{S{\'y}kora} J.,  {Ryb{\'a}k} J.,  2010, \mn@doi [\solphys]
  {10.1007/s11207-009-9483-x}, \href
  {https://ui.adsabs.harvard.edu/abs/2010SoPh..261..321S} {261, 321}

\bibitem[\protect\citeauthoryear{{Temmer}, {Ryb{\'a}k}, {Bend{\'\i}k},
  {Veronig}, {Vogler}, {Otruba}, {P{\"o}tzi}  \& {Hanslmeier}}{{Temmer}
  et~al.}{2006}]{Tetal06}
{Temmer} M.,  {Ryb{\'a}k} J.,  {Bend{\'\i}k} P.,  {Veronig} A.,  {Vogler} F.,
  {Otruba} W.,  {P{\"o}tzi} W.,   {Hanslmeier} A.,  2006, \mn@doi [\aap]
  {10.1051/0004-6361:20054060}, \href
  {https://ui.adsabs.harvard.edu/abs/2006A&A...447..735T} {447, 735}

\bibitem[\protect\citeauthoryear{{Tlatov}, {Kuzanyan}  \&
  {Vasil'yeva}}{{Tlatov} et~al.}{2016}]{Tetal16}
{Tlatov} A.~G.,  {Kuzanyan} K.~M.,   {Vasil'yeva} V.~V.,  2016, \mn@doi
  [\solphys] {10.1007/s11207-016-0880-7}, \href
  {https://ui.adsabs.harvard.edu/abs/2016SoPh..291.1115T} {291, 1115}

\bibitem[\protect\citeauthoryear{{Tobias}}{{Tobias}}{1997}]{T97}
{Tobias} S.~M.,  1997, \aap, \href
  {https://ui.adsabs.harvard.edu/abs/1997A&A...322.1007T} {322, 1007}

\bibitem[\protect\citeauthoryear{{Ulrich} \& {Tran}}{{Ulrich} \&
  {Tran}}{2013}]{UT13}
{Ulrich} R.~K.,  {Tran} T.,  2013, \mn@doi [\apj]
  {10.1088/0004-637X/768/2/189}, \href
  {https://ui.adsabs.harvard.edu/abs/2013ApJ...768..189U} {768, 189}

\bibitem[\protect\citeauthoryear{{Vecchio}, {Laurenza}, {Meduri}, {Carbone}  \&
  {Storini}}{{Vecchio} et~al.}{2012}]{Vetal12}
{Vecchio} A.,  {Laurenza} M.,  {Meduri} D.,  {Carbone} V.,   {Storini} M.,
  2012, \mn@doi [\apj] {10.1088/0004-637X/749/1/27}, \href
  {https://ui.adsabs.harvard.edu/abs/2012ApJ...749...27V} {749, 27}

\bibitem[\protect\citeauthoryear{{Vizoso} \& {Ballester}}{{Vizoso} \&
  {Ballester}}{1990}]{VB90}
{Vizoso} G.,  {Ballester} J.~L.,  1990, \aap, \href
  {https://ui.adsabs.harvard.edu/abs/1990A&A...229..540V} {229, 540}

\bibitem[\protect\citeauthoryear{{Weiss}}{{Weiss}}{2010}]{W10}
{Weiss} N.,  2010, \mn@doi [Astronomy and Geophysics]
  {10.1111/j.1468-4004.2010.51309.x}, \href
  {https://ui.adsabs.harvard.edu/abs/2010A&G....51c...9W} {51, 3.09}

\bibitem[\protect\citeauthoryear{{Wilson}}{{Wilson}}{1994}]{W94}
{Wilson} P.~R.,  1994, Cambridge Astrophysics Series, \href
  {https://ui.adsabs.harvard.edu/abs/1994CAS....24.....W} {24}

\bibitem[\protect\citeauthoryear{{Wilson}, {Altrocki}, {Harvey}, {Martin}  \&
  {Snodgrass}}{{Wilson} et~al.}{1988}]{Wetal88}
{Wilson} P.~R.,  {Altrocki} R.~C.,  {Harvey} K.~L.,  {Martin} S.~F.,
  {Snodgrass} H.~B.,  1988, \mn@doi [\nat] {10.1038/333748a0}, \href
  {https://ui.adsabs.harvard.edu/abs/1988Natur.333..748W} {333, 748}

\bibitem[\protect\citeauthoryear{{Zhang}, {Mursula}, {Usoskin}, {Wang}  \&
  {Du}}{{Zhang} et~al.}{2011}]{Zetal11}
{Zhang} L.,  {Mursula} K.,  {Usoskin} I.,  {Wang} H.,   {Du} Z.,  2011, in
  Astronomical Society of India Conference Series. pp 175--180

\makeatother
\end{thebibliography}

\end{document}